\documentclass[aps,prc,twocolumn,longbibliography,reprint,groupedaddress,nofootinbib]{revtex4-1} 
\usepackage{amsmath,amsfonts,amsthm,amssymb}
\usepackage[utf8]{inputenc}
\usepackage[dvips]{graphics,graphicx}
\usepackage[usenames,dvipsnames]{color}
\definecolor{darkblue}{RGB}{0,0,196}
\definecolor{darkgreen}{RGB}{0,120,0}
\usepackage[colorlinks=true,linktocpage=true,linkcolor=darkgreen,citecolor=darkgreen,
urlcolor=darkgreen]{hyperref}
\usepackage{ulem}
\usepackage{cancel}
\usepackage{multirow}
\usepackage{longtable}
\usepackage{textcomp}
\usepackage{natbib}

\begin{document}

\preprint{}

\title{Second order hydrodynamics based on effective kinetic theory and electromagnetic signals from QGP}

\author{Lakshmi J. Naik}
\email{jn\_lakshmi@cb.students.amrita.edu}
%
\author{V. Sreekanth}
\email{v\_sreekanth@cb.amrita.edu}
\affiliation{Department of Sciences, Amrita School of Physical Sciences, Coimbatore, Amrita Vishwa Vidyapeetham, India}
\date{\today}

\begin{abstract}
We study the thermal dilepton and photon production from relativistic heavy ion collisions in presence of viscosities by employing the recently developed  
second order dissipative hydrodynamic formulation estimated within a quasiparticle description of 
thermal QCD (Quantum Chromo-Dynamics) medium.
The sensitivity of shear and bulk viscous pressures to the temperature dependence of relaxation time is analyzed within one dimensional boost invariant expansion of quark gluon plasma (QGP).
The dissipative corrections
to the phase-space distribution functions upto first order in gradients are obtained from the Chapman-Enskog like iterative solution 
of effective Boltzmann equation in the relaxation time approximation. Thermal dilepton and photon 
production rates for QGP are calculated by employing this viscous modified distribution function. Yields of these particles are quantified for the longitudinal expansion of QGP with different temperature dependent relaxation times. 
Our analysis employing this second order hydrodynamic model indicates 
that the spectra of dileptons and photons gets enhanced by both bulk and shear viscosities and is well behaved. 
Also, these particle yields are found to be sensitive to relaxation time. Further, we do a comparison of these particle spectra with a standard hydrodynamic formulation.

\end{abstract}

\maketitle

\section{Introduction}
\label{intro}

The experiments at Relativistic Heavy Ion Collider (RHIC) and at Large Hadron 
Collider (LHC) suggest the existence of strongly coupled quark-gluon matter at
extreme high 
temperature and density~\cite{STAR:2005gfr,PHENIX:2004vcz,BRAHMS:2004adc,PHOBOS:2004zne}. These experiments provide opportunity to inspect the 
features of the hot nuclear matter, QGP which is believed to
have existed in the 
primordial universe. 
Theoretical and experimental investigations of properties 
of the hot and dense QGP is being pursued vigorously by the community~\cite{Jaiswal:2020hvk}. 
Analysis of the experimental data from the heavy-ion collision experiments imply that the QGP 
has a near perfect fluid nature with 
extreme low value of shear viscosity to entropy ratio, $\eta/s = 1/4\pi$~\cite{Kovtun:2004de}.
This surprisingly low value of viscosity has evoked interest in the 
application of relativistic dissipative hydrodynamics to heavy ion collisions~\cite{Romatschke:2017ejr}. 

The evolution of viscous QGP has to be modelled using higher order relativistic viscous 
hydrodynamical theories, since first order Navier-Stokes theory exhibits acausal 
behaviour and numerical instabilities~\cite{Hiscock:1983zz,Hiscock:1985zz,Geroch:1990bw}. Recently, several investigations are being pursued towards the development of causal first order hydrodynamic theory~\cite{Bemfica:2017wps,Hoult:2020eho,Biswas:2022cla}. 
When it comes to second order theories, there is no unique prescription 
and there exist several successful formalisms. The earliest attempts 
in this direction were by Muller, Israel and Stewart~\cite{Muller1967329,Israel:1979wp}. 
Development of second order causal 
theories is an active field of research and several new formalisms have been 
proposed and studied in the context of heavy ion collisions~\cite{Jaiswal:2016hex,Romatschke:2017ejr}. 
Second order viscous hydrodynamics developed in Ref.~\cite{Bhadury:2020ivo} is a recently 
proposed formalism within the effective fugacity quasiparticle model (EQPM) 
for the hot QCD medium~\cite{Chandra:2009jjo,Chandra:2011en}. 
The EQPM prescription incorporates the thermal QCD medium interaction effects into a system 
of quasipartons by considering thermal modifications to the phase-space distribution 
functions in terms of effective quark and gluon fugacities. The dissipative hydrodynamic evolution equations
have been determined by employing the effective covariant kinetic theory developed for EQPM~\cite{Mitra:2018akk}.
The non-equilibrium corrections to the distribution functions have been estimated 
in this hydrodynamic framework by the Chapman-Enskog (CE) expansion in relaxation 
time approximation (RTA)~\cite{Jaiswal:2013npa,Jaiswal:2013vta}. Moreover, a quasiparticle model such as EQPM distinguishes the quark and gluon sector in the hot QCD medium and can be used to study thermal dilepton and photon emission from QGP~\cite{Chandra:2015rdz}.

Properties of the QCD matter created in high energy heavy ion
collisions can be 
studied by analysing
the different signals emitted, such as thermal dileptons and photons.  
As these particles interact only electromagnetically,  
they can easily decouple from the strong nuclear matter and reach the detectors
without further interaction with other particles.
Effect of viscosities has consequences on the 
thermal dilepton and photon spectra from heavy ion
collisions. 
The role of shear viscosity on
thermal particle 
production using causal dissipative hydrodynamics 
has been investigated in Refs.~\cite{Dusling:2008xj,Dusling:2009bc,Bhatt:2009zg,Bhatt:2011kx,Chaudhuri:2011up,Vujanovic:2013jpa,Vujanovic:2017psb}.
The influence of bulk viscosity on the
expansion of QGP and signals emanating from it has been studied in Refs.~\cite{Torrieri:2007fb,Fries:2008ts,Rajagopal:2009yw,Bhatt:2010cy,Bhatt:2011kx,Paquet:2015lta}. 
In certain situations, the inclusion of dissipation 
into the heavy ion collision scenario may induce a
phenomenon called \textit{cavitation} which causes the 
hydrodynamic description to be invalid before the 
freeze-out and thereby affecting the signals~\cite{Rajagopal:2009yw,Bhatt:2011kr}. 
Recently, thermal particle production
has been investigated by employing the concept
of hydrodynamic attractors for temperature evolution~\cite{Coquet:2021lca,Naik:2021yph,Naik:2021yue}. The EQPM prescription has been  employed to study several observables from heavy ion collisions. Impact of chromo-Weibel instability on thermal dilepton emission has been analysed by the authors of Ref.~\cite{Chandra:2016dwy} within EQPM.
 The effect due to collisional contributions of thermal QCD medium has also been investigated on thermal dileptons~\cite{Naik:2020jfc} and heavy quark transport~\cite{Prakash:2021lwt}, using EQPM.
However, in these studies employing the EQPM, ideal hydrodynamics was used to model the evolution of the system.

In the present analysis,
we proceed
to implement the causal second order hydrodynamic framework within EQPM
to study the evolution of QGP and
thermal particle production from heavy ion collisions. 
We analyze the evolution for different temperature dependent relaxation times. 
The thermal particle emission rates have
to be calculated by proper modelling of the momentum distribution
functions incorporating the viscous effects. The non-equilibrium part of distribution 
function is generally determined by the 14-moment Grad's method
or the CE expansion, out of these, the CE type expansion of non-equilibrium 
distribution function is shown to be well behaved even up to second 
order in gradients~\cite{Bhalerao:2013pza}. In this work, we intend 
to calculate thermal dilepton and photon production rates 
using the 
viscous modified quasiparticle thermal distribution functions
determined by the CE method in RTA.
The particle emission spectra is dependent on the temperature profile of QGP and 
is found by modelling the expansion of QGP with 
relativistic dissipative hydrodynamics under one-dimensional ($1-$D) boost invariant Bj\"{o}rken expansion. 

The paper is structured as follows. In section II, we review 
the EQPM description and the formalism used to derive the second order viscous hydrodynamics. 
The causal hydrodynamic evolution equations of the expanding QGP medium are prescribed within the 1-D boost invariant Bj\"orken flow. 
Then, we analyze the evolution of shear stress and bulk viscous pressure for different relaxation times. Section III is devoted to the
calculation of thermal dilepton and photon production rates in
the presence of 
viscous modified momentum distribution function. 
In section IV, thermal particle 
spectra is computed in 1-D Bj\"{o}rken expansion and we present the results of our analysis
in section V. Section VI details conclusions and future outlook.

\textbf{Notations and conventions:} We are working with natural units $i.e.,\,c=\hbar=k_B=1$.
The 
Minkowski metric is taken as $\eta^{\mu\nu} = \textrm{diag}(1,-1,-1,-1)$. The fluid four-velocity is denoted as $u^\mu$, which is normalized as $u^\mu u_\mu =1$ and in the local rest frame (LRF) of the fluid, $u^\mu = (1,0,0,0)$.
The quantity $\Delta^{\mu\nu} = \eta^{\mu\nu} - u^\mu u^\nu$ represents the projection
operator orthogonal to $u^\mu$ and $\nabla^\mu = \Delta^{\mu\nu} \partial_\nu$.
Also, $\Delta_{\alpha\beta}^{\mu\nu} \equiv \frac{1}{2}(\Delta_\alpha^\mu \Delta_\beta^\nu + 
\Delta_\beta^\mu \Delta_\alpha^\nu) - \frac{1}{3}\Delta^{\mu\nu} \Delta_{\alpha\beta}$ is the 
traceless symmetric projection operator orthogonal to $u^\mu$.

\section{Second order dissipative hydrodynamics based on quasiparticle model}
\label{eqpm_hydro}
In this section, we present the formalism to estimate the second order relativistic viscous hydrodynamic equations within the EQPM description of QCD medium as derived in Ref.~\cite{Bhadury:2020ivo}. 
%
The EQPM maps the thermal QCD medium effects through temperature dependent effective fugacity parameters of constituent noninteracting quasiparticles.
Within this model, the equilibrium momentum distribution functions of the quasiparticles are given by~\cite{Chandra:2009jjo},
\begin{equation} \label{QPM_f0}
 f_k^0 = \frac{z_k \exp[-\beta(u_\mu p_k^\mu)]}{1 \pm z_k \exp[-\beta(u_\mu p_k^\mu)]}.
\end{equation}
Here, $k \equiv (q,g)$ represent the quarks and gluons respectively and $z_q, z_g$ denote the effective quark, gluon fugacities which capture the QCD medium interactions. The model divides the hot QCD medium into two sectors : (i) the effective gluonic sector, which describes the contribution of gluonic action to the pressure as well as the contribution due to the internal fermion lines, (ii) the matter sector, which represents the interactions between quarks and antiquarks, along with their interactions with gluons.
The form of $z_g$ and $z_q$ is determined by equating the expression for pressure obtained within EQPM with the lattice data (see Ref.~\cite{Chandra:2011en} for detailed discussion). Here, the ($2+1$) 
flavor lattice
QCD equation of state~\cite{Cheng:2007jq,Borsanyi:2013bia} is considered. 
In Eq.~\eqref{QPM_f0}, $p_k^\mu = (E_k, \vec{p}_k)$ represent the particle four-momenta of quarks and gluons and 
$\beta = 1/T$ is the inverse of temperature. The quasiparticle momenta ($\tilde{p}_k^\mu$)
and bare particle momenta ($p_k^\mu$) are related through the dispersion relation:
\begin{equation}  \label{DIS_EQ}
 \tilde{p}_{g,q}^\mu = p_{g,q}^\mu + \delta \omega_{g,q} u^\mu;\,\,\,\,\,\,\,\,\,\,\,
 \delta \omega_{g,q} = T^2 \partial_T \ln (z_{g,q}),
\end{equation}
where $\delta \omega_{g,q}$ denotes the modified part of the dispersion relation. From Eq.~\eqref{DIS_EQ}, the single particle energy of the quasiparticles is given by 
\begin{equation}  \label{E_modified}
\tilde{p}_{g,q}^0 \equiv  \omega_{g,q} = E_{g,q} + \delta \omega_{g,q}.
\end{equation}

In order to obtain the viscous hydrodynamic equations, one need to quantify the dissipative 
corrections of the system. 
The non-equilibrium corrections to the phase space distribution
functions are estimated by considering the effective Boltzmann equation within EQPM. 
The form of relativistic transport equation in RTA for the collision term is obtained as~\cite{Mitra:2018akk}
\begin{equation} \label{BE_RTA}
 \tilde{p}_k^\mu \partial_\mu f_k^0(x,\tilde{p}_k) + F_k^\mu \partial_\mu^{(p)} f_k^0
 = -\frac{\delta f_k}{\tau_R} \omega_k,
\end{equation}
where $\tau_R$ is the relaxation time and $\delta f_k$ denotes the non-equilibrium part of the distribution
function. The force term is defined from energy-momentum and particle flow conservation and
is given by $F_k^\mu = - \partial_\nu (\delta \omega_k u^\nu u^\mu)$~\cite{Mitra:2018akk}.

Using the Landau definition : $u_\nu T^{\mu\nu} = \epsilon u^\mu$~\cite{Landau:1987}, the energy-momentum tensor can be decomposed in terms of hydrodynamic degrees of freedom as
\begin{equation} \label{T_hydro}
 T^{\mu\nu} = \epsilon u^\mu u^\nu - P\Delta^{\mu\nu} + \tau^{\mu\nu},
\end{equation}
where $\epsilon$ and $P$ denote the equilibrium energy density and pressure of the system respectively.
$\tau^{\mu\nu}=\pi^{\mu\nu} - \Pi \Delta^{\mu\nu}$ is the dissipative current, with $\pi^{\mu\nu}$ being the traceless part.
The quantities $\pi^{\mu\nu}$ and $\Pi$ represent the shear stress tensor and bulk viscous pressure respectively.
The evolution equations for $\epsilon$ and 
$u^\mu$ can be obtained by projecting the energy-momentum conservation equation
$\partial_{\mu} T^{\mu\nu} = 0$ along and orthogonal to $u^\mu$ and are given by
\begin{eqnarray}
 \dot{\epsilon} + (\epsilon + P + \Pi) \Theta - \pi^{\mu\nu}\sigma_{\mu\nu} &=& 0, \\
 (\epsilon + P + \Pi) \dot{u}^\alpha - \nabla^\alpha (P + \Pi) + 
 \Delta_\nu^\alpha \partial_\mu \pi^{\mu\nu} &=& 0.
\end{eqnarray}
Here, $\Theta \equiv \partial^\mu u_\mu$ is the four-divergence of fluid velocity, $\sigma^{\mu\nu} \equiv \Delta_{\alpha\beta}^{\mu\nu} \nabla^\alpha u^\beta$ represents the
shear stress tensor and  $\dot{A} \equiv u^\mu \partial_\mu A$ is the co-moving derivative of $A$. The derivatives of $\beta$ can be calculated from the above equations
\begin{eqnarray}
 \dot{\beta} &=& \beta c_s^2 \left(\Theta + \frac{\Pi \Theta - \pi^{\mu\nu}\sigma_{\mu\nu}}
 {\epsilon + P}\right),  \\
 \nabla^\alpha \beta &=& -\beta \left(\dot{u}^\alpha + 
 \frac{\Pi \dot{u}^\alpha - \nabla^\alpha \Pi + \Delta_\nu^\alpha \partial_\mu \pi^{\mu\nu}}
 {\epsilon + P}\right),
\end{eqnarray}
where $c_s^2 = dP/d\epsilon$ denotes the speed of sound squared.

Now, the form of energy-momentum tensor in terms of quasiparticle 
four-momenta can be written as
\begin{eqnarray} \label{T_eqpm}
 T^{\mu\nu} (x) &=& \sum_k g_k \int d\tilde{P}_k \tilde{p}_k^\mu \tilde{p}_k^\nu f_k(x,\tilde{p}_k) 
 \nonumber\\
 &&+ \sum_k g_k \delta \omega_k \int d\tilde{P}_k 
 \frac{< \tilde{p}_k^\mu \tilde{p}_k^\nu >}{E_k} f_k(x,\tilde{p}_k),
\end{eqnarray}
with $g_k$ being the degeneracy factor of quasiparticles. Here, $< \tilde{p}_k^\mu \tilde{p}_k^\nu > \equiv \frac{1}{2}(\Delta_\alpha^\mu \Delta_\beta^\nu + \Delta_\beta^\mu \Delta_\alpha^\nu)\tilde{p}_k^\alpha \tilde{p}_k^\beta$ and $d\tilde{P}_k \equiv \frac{d^3|\vec{p}_k|}{(2\pi)^3\omega_k}$ represents the phase-space factor.
Using Eqs.~\eqref{T_hydro} and \eqref{T_eqpm}, the expressions for shear stress tensor and bulk viscous pressure are obtained as~\cite{Mitra:2018akk} 
\begin{eqnarray}
 \pi^{\mu\nu} &=& \sum_k g_k \Delta_{\alpha\beta}^{\mu\nu} \int d\tilde{P}_k 
 \tilde{p}_k^\alpha \tilde{p}_k^\beta \delta f_k  \nonumber\\
 &&+ \sum_k g_k \delta \omega_k \Delta_{\alpha\beta}^{\mu\nu} \int d\tilde{P}_k 
 \tilde{p}_k^\alpha \tilde{p}_k^\beta \frac{\delta f_k}{E_k}, \label{pi_eqn} \\
 \Pi &=& -\frac{1}{3}\sum_k g_k \Delta_{\alpha\beta}\int d\tilde{P}_k 
 \tilde{p}_k^\alpha \tilde{p}_k^\beta \delta f_k  \nonumber\\
 &&-\frac{1}{3} \sum_k g_k \delta \omega_k \Delta_{\alpha\beta} \int d\tilde{P}_k 
 \tilde{p}_k^\alpha \tilde{p}_k^\beta \frac{\delta f_k}{E_k}.   \label{Pi_eqn}
\end{eqnarray}
The transport coefficients of the system can be determined once we know the form of $\delta f_k$. 
In Ref.~\cite{Bhadury:2020ivo}, viscous correction is calculated by employing the Chapman-Enskog like iterative solution of the Boltzmann equation in 
RTA.
Within the effective kinetic theory considered, the 
non-equilibrium corrections to the quark (antiquark)
distribution function up to first order in gradients is then obtained as
\begin{eqnarray} \label{delta_f1}
 \delta f_q &=& \tau_R \bigg[ \tilde{p}_q^\mu\partial_\mu \beta 
 + \frac{\beta\, \tilde{p}_q^\mu\, \tilde{p}_q^\nu}{u\!\cdot\!\tilde{p}_q}\partial_\mu u_\nu 
 - \beta \Theta (\delta\omega_q) \nonumber\\
 &&- \beta \dot{\beta} \left(\frac{\partial(\delta\omega_q)}{\partial\beta}\right)\bigg] 
 f_{q}^0\bar{f}_{q}^0, 
\end{eqnarray}
where $\bar{f}_{q}^0 = 1 - f_{q}^0$. By employing the above equation in Eqs.~\eqref{pi_eqn} 
and \eqref{Pi_eqn}, and considering $\tau_R$ to be independent of momenta, the following relations are obtained:
\begin{eqnarray} \label{I_order_ev}
 \pi^{\mu\nu} = 2 \tau_R \beta_\pi \sigma^{\mu\nu},  
 \,\,\,\,\,\,\,\,\,\,
 \Pi = -\tau_R \beta_\Pi \Theta.
\end{eqnarray}
We note that, as a result of RTA, there will be only a single time-scale for both shear and bulk relaxation times.
First order transport coefficients within the effective covariant kinetic theory can be 
determined by comparing the above equations with that of the Navier-Stokes
equation
\begin{equation}
 \eta = \tau_R \beta_\pi, \,\,\,\,\,\,\,\,\, \zeta = \tau_R \beta_\Pi.
\end{equation}
Here, $\eta$ and $\zeta$ are the coefficients of shear and bulk viscosities respectively. 
The quantities $\beta_\pi, \beta_\Pi$ appearing in the above equations are determined in terms
of thermodynamic integrals and are given as
\begin{eqnarray}
 \beta_\pi &=& \beta\sum_k \bigg[\Tilde{J}^{(1)}_{k~42} 
 + \delta\omega_k\Tilde{L}^{(1)}_{k~42} \bigg],\label{beta pi}\\
\beta_\Pi &=& \beta \sum_k \bigg[c_s^2  \bigg(\Tilde{J}_{k~31}^{(0)} + 
\delta\omega_k \Tilde{L}_{k~31}^{(0)}   \nonumber\\
&&-\frac{\partial(\delta\omega_k)}{\partial\beta} 
\left(\Tilde{J}_{k~21}^{(0)} + \delta\omega_k \Tilde{L}_{k~21}^{(0)}\right)\bigg) \nonumber\\
&&+ \frac{5}{3}\bigg(\Tilde{J}_{k~42}^{(1)} 
+ \delta\omega_k \Tilde{L}_{k~42}^{(1)} \bigg)      
- \delta\omega_k \left(\Tilde{J}_{k~21}^{(0)} 
+ \delta\omega_k \Tilde{L}_{k~21}^{(0)}\right) \bigg]\label{beta Pi}. \nonumber\\
\end{eqnarray}
The form of thermodynamic integrals, $\Tilde{J}^{(r)}_{k~nm}$ and $\Tilde{L}^{(r)}_{k~nm}$ appearing in the above equations are given in 
Appendix~\ref{A}.
Following the methodology of
Ref.~\cite{Bhadury:2020ivo}, the evolution equations for shear stress tensor and 
bulk viscous pressure are obtained as
\begin{eqnarray} \label{shear evolution def hydro}
\dot{\pi}^{\langle\mu\nu\rangle} + \frac{\pi^{\mu\nu}}{\tau_R} &=& 2 \beta_{\pi} \sigma^{\mu\nu} 
+ 2 \pi_{\phi}^{\langle\mu} \omega^{\nu\rangle\phi} 
- \delta_{\pi\pi} \pi^{\mu\nu} \theta \nonumber\\
&&- \tau_{\pi\pi} \pi_{\phi}^{\langle\mu} \sigma^{\nu\rangle \phi} 
+ \lambda_{\pi\Pi} \Pi \sigma^{\mu\nu},\\
\dot{\Pi} + \frac{\Pi}{\tau_R} &=& -\beta_{\Pi} \theta - \delta_{\Pi\Pi} \Pi \theta 
+ \lambda_{\Pi\pi} \pi^{\mu\nu} \sigma_{\mu\nu}.\label{bulk evolution def hydro}
\end{eqnarray}
Here, $\omega^{\mu\nu} = \frac{1}{2}(\nabla^\mu u^\nu - \nabla^\nu u^\mu)$
denotes the vorticity tensor.
The second order transport coefficients in Eqs.~\eqref{shear evolution def hydro} and
\eqref{bulk evolution def hydro} are obtained in terms of 
different thermodynamic integrals and are shown in Appendix~\ref{A}.


\subsection{Viscous evolution for different relaxation times
}
\label{Bjorken}

We solve the second order viscous evolution equations, Eqs.~\eqref{shear evolution def hydro} and
\eqref{bulk evolution def hydro}, by choosing different temperature dependent forms for the relaxation time. We model the expansion using the 1-D Bj\"{o}rken flow~\cite{Bjorken:1982qr}, which considers the 
QGP medium as a transversely homogenous and longitudinally boost-invariant system. It is now convenient to parameterize the coordinates as $t = \tau\cosh\eta_s$ and $z=\tau\sinh\eta_s$, where 
$\tau = \sqrt{t^2 -z^2}$ is the proper time and $\eta_s = \frac{1}{2}\ln\frac{t+z}{t-z}$
is the space-time rapidity of the system. Fluid four-velocity
is expressed using the ansatz, $u^\mu =(\cosh\eta_s,0,0,\sinh\eta_s)$.
Now, within this model, the 
hydrodynamic equations given by Eqs.~\eqref{shear evolution def hydro} 
and~\eqref{bulk evolution def hydro} 
reduce to the following coupled non-linear differential equations in $\tau$~\cite{Bhadury:2020ivo}:
\begin{eqnarray}    \label{energy_evolution}
 \frac{d \epsilon}{d\tau} &=& -\frac{1}{\tau}\left(\epsilon + P + \Pi - \pi \right),    \\
 \frac{d \pi}{d\tau} + \frac{\pi }{\tau_\pi} &=& \frac{4}{3}\frac{\beta_\pi}{\tau} 
 -\left( \frac{1}{3}\tau_{\pi\pi} + \delta_{\pi\pi}\right) \frac{\pi}{\tau} 
 + \frac{2}{3} \lambda_{\pi \Pi} \frac{\Pi}{\tau},  \label{pi_evolution}\\
 \frac{d \Pi}{d\tau} + \frac{\Pi }{\tau_\Pi} &=& -\frac{\beta_\Pi}{\tau} 
 -\delta_{\Pi\Pi} \frac{\Pi}{\tau} + \lambda_{\Pi \pi} \frac{\pi}{\tau};   \label{Pi_evolution}
\end{eqnarray}
where $\pi = \pi^{00}-\pi^{zz}$. 
We note that, under Bj\"{o}rken expansion, $\omega^{\mu\nu} = 0$ which implies that the term 
$\pi_{\phi}^{\langle\mu} \omega^{\nu\rangle\phi}$ in Eq.~\eqref{shear evolution def hydro}
vanishes and has no impact on the evolution of QGP. Temperature dependence of the shear
and bulk second order transport coefficients appearing in the above equations is analyzed in 
Ref.~\cite{Bhadury:2020ivo} and their analysis indicate that the thermal QCD medium effects
have significant impact on these coefficients. 
Note that,
Eqs.~\eqref{pi_evolution} and \eqref{Pi_evolution} give the evolution for $\pi$ and $\Pi$ respectively governed by 
their corresponding relaxation times. 
These equations, coupled with Eq.~\eqref{energy_evolution} can be solved numerically by fixing $\tau_\pi, \tau_\Pi$. 
In the present work, in order to study the effect of relaxation time on evolution and subsequently on signals, we choose different temperature dependent values lying between $\tau_R = \frac{2(2-\ln 2)}{T}(\eta/s)$, the result corresponding to $N=4$ supersymmetric Yang-Mills theory~\cite{Baier:2007ix,Natsuume:2007ty} and $\tau_R=\frac{5.0...5.9}{T}(\eta/s)$, which is motivated by kinetic theory~\cite{York:2008rr}. 
We take the following forms of relaxation time in our analysis : $\tau_\pi = b_1 (\eta/s)/T$ and $\tau_\Pi = b_2 (\zeta/s)/T$. The value of $\eta/s$ is taken to be $ 1/4\pi$~\cite{Kovtun:2004de} and $\zeta/s$ as temperature dependent according to the studies of strongly interacting gauge theories as given below~\cite{Kanitscheider:2009as}
\begin{equation}\label{zetabys}
    \frac{\zeta}{s} = 2\frac{\eta}{s}\left( \frac{1}{3} - c_s^2 \right) \equiv \kappa(T) \frac{\eta}{s}.
\end{equation}
As mentioned earlier, due to RTA, we obtain a single 
relaxation time-scale for both shear and bulk $i.e., \tau_\pi = \tau_\Pi = \tau_R$. This condition together with Eq.~\eqref{zetabys} would give the relation $b_2 = b_1/\kappa(T)$, which ensures that shear and bulk relaxation times 
are equal for any value of $b_1$. 
Following the analysis in Ref.~\cite{Bhadury:2020ivo}, we also consider a constant temperature independent relaxation time, $\tau_R = 0.15$ fm/c in our study.

\begin{figure}[t!]
 \begin{center}
  \includegraphics[width=\linewidth]{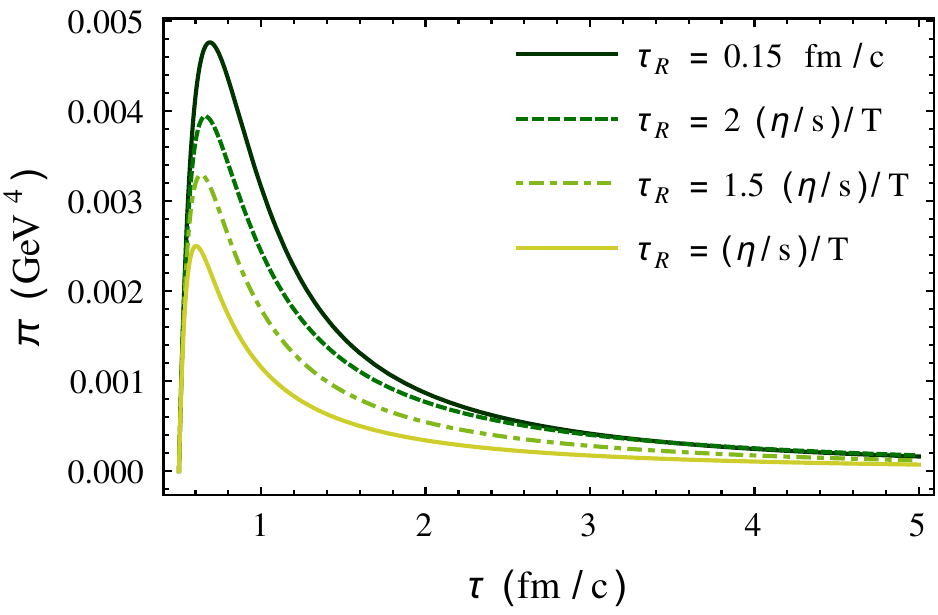}
 \end{center}
 \vspace{-0.8cm}
 \caption{Proper time evolution of shear stress tensor for different temperature dependent forms of $\tau_R$. Initial conditions are taken to be $T_0=0.31$ GeV and $\tau_0 = 0.5$ fm/c.
 Evolution corresponding to $\tau_R = 0.15$ fm/c is also plotted for comparison.}
 \label{shear_evo}
\end{figure}

\begin{figure}[t!]
 \begin{center}
  \includegraphics[width=\linewidth]{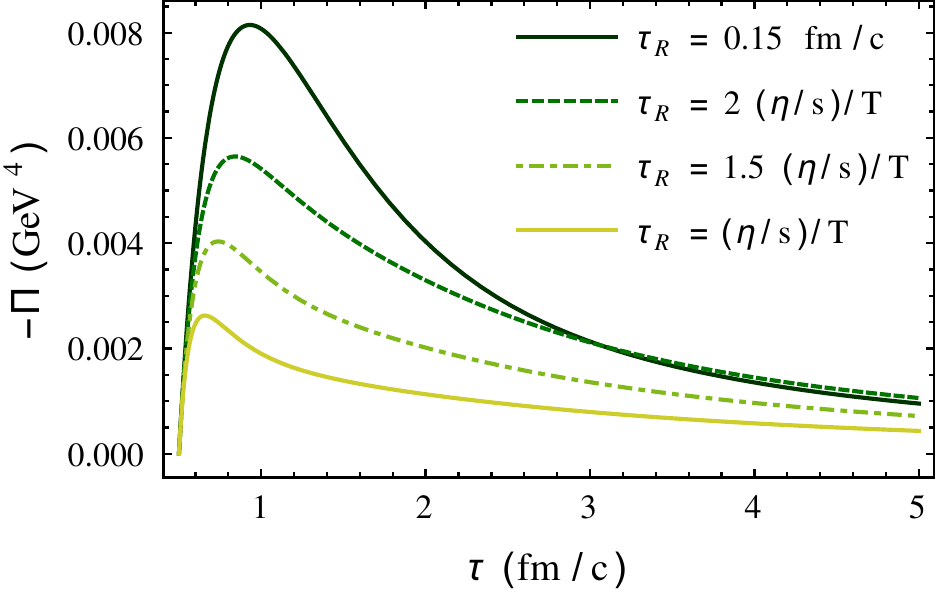}
 \end{center}
 \vspace{-0.8cm}
 \caption{Proper time evolution of bulk viscous pressure for different temperature dependent forms of $\tau_R$. Initial conditions are taken to be $T_0=0.31$ GeV and $\tau_0 = 0.5$ fm/c. 
 Evolution corresponding to $\tau_R = 0.15$ fm/c is also plotted for comparison.}
 \label{bulk_evo}
\end{figure}

Now, we investigate the sensitivity of dissipative quantities to the temperature dependence of $\tau_R$.
We solve Eqs.~\eqref{energy_evolution}, \eqref{pi_evolution} and \eqref{Pi_evolution} numerically by providing the initial conditions relevant to RHIC energies.
We choose the initial
proper time and temperature to be $\tau_0=0.5$ fm/c and $T_0 = 0.31$ GeV respectively, following the previous studies~\cite{Srivastava:1999ekv,Bhatt:2010cy,Chandra:2015rdz}.
The initial values of viscous contributions are taken as
$\pi(\tau_0)= \Pi(\tau_0)= 0$ GeV/fm$^3$. The (2+1) flavor lattice QCD EoS~\cite{Cheng:2007jq,Borsanyi:2013bia} is used to close this system of equations.
In Figs.~\ref{shear_evo} and \ref{bulk_evo}, the proper time
evolution of shear stress tensor and magnitude of bulk viscous pressure respectively are plotted for different temperature dependent $\tau_R$. 
The evolution is plotted for $\tau_R = 2(\eta/s)/T,\,1.5(\eta/s)/T$ and $(\eta/s)/T$. We observe that both shear and bulk pressures have strong dependency on the form of $\tau_R$. The effect of viscous contributions are high for $\tau_R = 2(\eta/s)/T$, while it is the lowest for $\tau_R = (\eta/s)/T$. We also plot the viscous terms for a constant value of relaxation time, $\tau_R = 0.15$ fm/c, chosen arbitrarily. 
We note that, at early times, the shear and bulk pressures are found to be high compared to the later times for every $\tau_R$ considered here.


\begin{figure}[t!]
 \begin{center}
  \includegraphics[width=\linewidth]{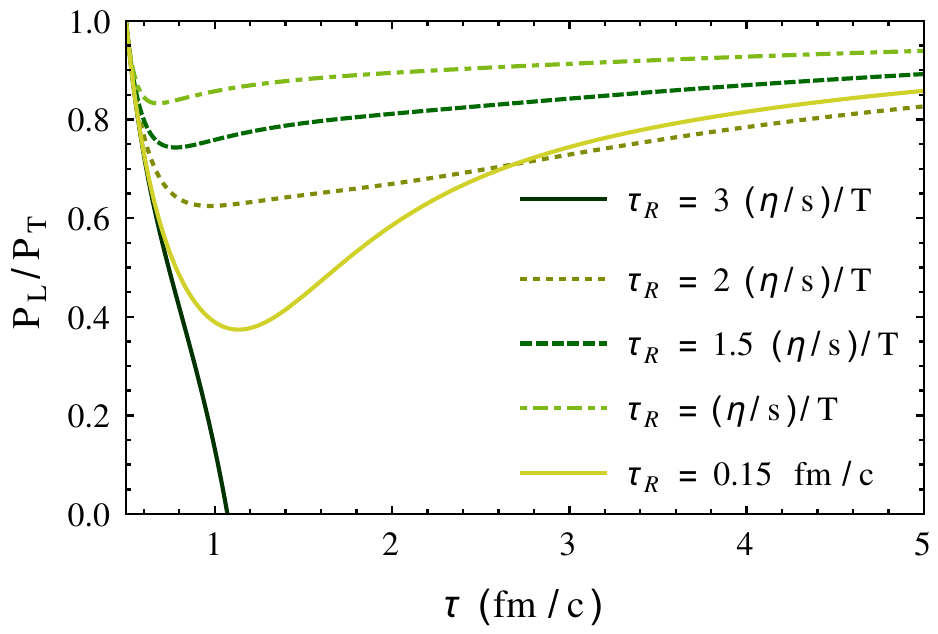}
 \end{center}
 \vspace{-0.8cm}
 \caption{Proper time evolution of pressure anisotropy $P_L/P_T$ with various temperature dependent relaxation times. Initial conditions are taken to be $T_0=0.31$ GeV and $\tau_0 = 0.5$ fm/c. 
 Evolution for constant $\tau_R$ is also shown for comparison.}
 \label{pLpT}
\end{figure}

In Fig.~\ref{pLpT}, we analyze the pressure anisotropy of the medium, which is defined as
\begin{equation}
    P_L/P_T  \equiv \frac{P+\Pi - \pi}{P+\Pi +\pi/2},
\end{equation}
for the various temperature forms of $\tau_R$. 
Here $P_L, P_T$ refer to the longitudinal, transverse pressures respectively. It can be seen that, as $\tau_R$ increases to $3(\eta/s)/T$, the ratio $P_L/P_T$ becomes negative causing cavitation in the medium and this stops the validity of hydrodynamics.
For $\tau_R = 3(\eta/s)/T$, we plot the curve only till pressure anisotropy approaches zero and it is observed that this occurs within a very small time of $1$ fm/c. In the same figure, we plot $P_L/P_T$ for the constant value, $\tau_R = 0.15$ fm/c and it is observed that cavitation scenario is not present for this particular case for the initial condition considered. However, cavitation can arise in the system even for very small values of temperature independent relaxation times greater than $0.15$ fm/c. For instance, we can observe cavitation for $\tau_R = 0.25$ fm/c (value taken in Ref.~\cite{Bhadury:2020ivo}), around $0.$ fm/c for the initial conditions taken in this work. 
\begin{figure}[t!]
 \begin{center}
  \includegraphics[width=\linewidth]{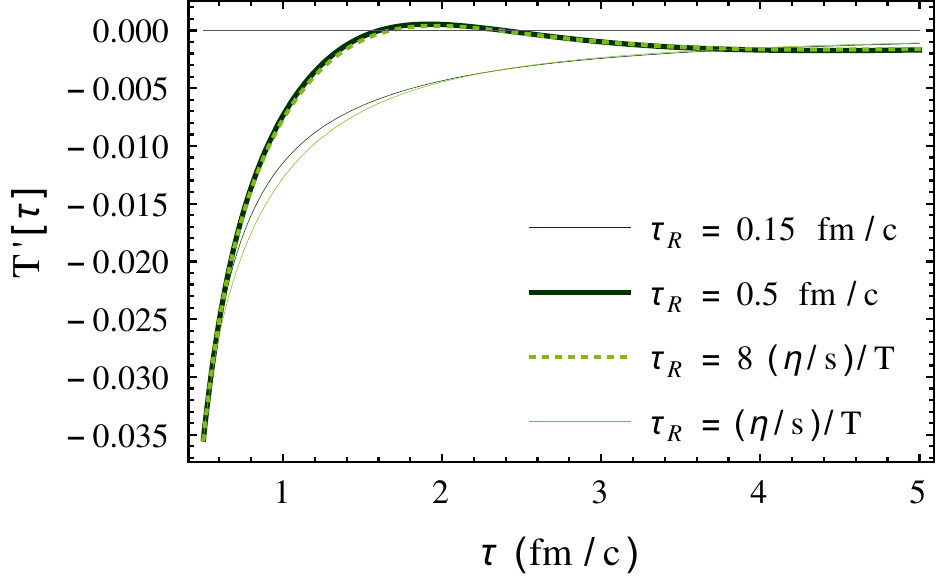}
 \end{center}
 \vspace{-0.8cm}
 \caption{Proper time evolution of $dT/d\tau$ by varying the relaxation time.}
 \label{dT}
\end{figure}

We now turn our attention towards the 
limiting values of relaxation times possible in such an 
application of the formalism to heavy ion collisions 
by looking to the rate of temperature variation of the fireball as it evolves. 
We plot the rate of change of temperature as a function of proper time for different 
$\tau_R$ as shown in Fig.~\ref{dT}. We observe that the slope of temperature evolution crosses the line $T'[\tau]=0$
as we increase the value of $\tau_R$. This indicates that, when $\tau_R$ is large, the system undergoes 
reheating as it expands. Reheating of the fireball is an unphysical scenario in relativistic heavy ion collisions, since it is expected that the temperature of QGP has to decrease monotonically during its expansion~\cite{Muronga:2001zk,Baier:2006um}.
This reheating can be observed for $\tau_R > 8(\eta/s)/T$ and for the constant values greater than $0.5$ fm/c.
It can be seen that, for $\tau_R = 8(\eta/s)/T$ and $0.5$ fm/c, $T'[\tau]$ crosses zero around $\tau = 2$ fm/c.

\begin{figure}[t!]
 \begin{center}
  \includegraphics[width=\linewidth]{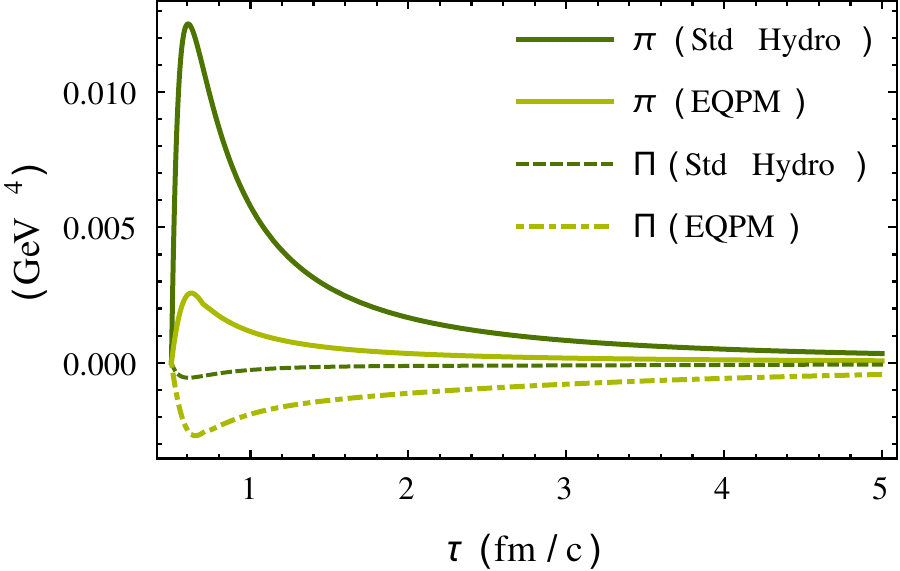}
 \end{center}
 \vspace{-0.8cm}
 \caption{Comparison of evolution of shear and bulk viscous pressures obtained within EQPM with that obtained using the standard hydrodynamics taken from Ref.~\cite{Muronga:2003ta}.}
 \label{evo_compare}
\end{figure}
Next, we compare the evolution of viscous quantities obtained within EQPM with another standard hydrodynamic framework. We take the second order dissipative hydrodynamics described in Ref.~\cite{Muronga:2003ta} for this analysis. Within this framework, the evolution equations for viscous quantities ($\pi$ and $\Pi$) in Bj\"orken flow are given by
\begin{eqnarray}
\frac{d\pi}{d\tau}&=&-\frac{\pi}{\tau_\pi} - \frac{\pi}{2}\left(\frac{1}{\tau} + \frac{1}{\beta_2}T\frac{d}{d\tau}\left(\frac{\beta_2}{T}\right)\right) + \frac{2}{3}\frac{1}{\beta_2}\frac{1}{\tau},\label{shear_std} \\
\frac{d\Pi}{d\tau}&=&-\frac{\Pi}{\tau_\Pi} - \frac{1}{2}\frac{\Pi}{\beta_0} \left(\frac{\beta_0}{\tau} + T\frac{d}{d\tau}\left(\frac{\beta_0}{T}\right)\right) \frac{1}{\beta_0}\frac{1}{\tau} \label{bulk_std},
\end{eqnarray}
where $\beta_0$ and $\beta_2$ are related to the relaxation times as $\tau_\Pi = \zeta \beta_0$ and $\tau_\pi = 2\eta\beta_2$. Note that, in our analysis, we have $\tau_\pi = \tau_\Pi = \tau_R$.
In Fig.~\ref{evo_compare}, we compare the evolution of shear and bulk viscous pressures within EQPM with that obtained using the above hydrodynamic framework. Equations \eqref{shear_std} and \eqref{bulk_std} are numerically solved together with the evolution equation for energy density (Eq.~\eqref{energy_evolution}) by providing the lattice EoS~\cite{Cheng:2007jq,Borsanyi:2013bia}. 
We adopt the same initial conditions as before and also take $\tau_R = (\eta/s)/T$ in our calculations.
The shear stress evolution in standard hydrodynamics is observed to be large compared to that of EQPM; while, the evolution of bulk pressure is high in the EQPM framework. Moreover, we note that for $\tau_R = 3(\eta/s)/T$, the pressure anisotropy does not become negative within this hydrodynamic description compared to the case of EQPM.
\section{Thermal dilepton and photon production rates}
\label{particle_rate}

Dissipation in the system influences the particle production rates in two ways:
firstly through the hydrodynamic evolution of the system and secondly via non-equilibrium
corrections to the single particle distribution functions. In the previous section,
we have analyzed the impact of viscosity on the evolution of QGP. Now, we incorporate the effect of viscosities on thermal dilepton and photon production rates through viscous modified distribution functions upto first order in gradients.
The major source of thermal dileptons in QGP medium is 
from the \textit{$q \bar{q}$-annihilation}, 
$q \bar{q} \rightarrow  \gamma^{*} \rightarrow l^{+} l^{-}$. 
From relativistic kinetic theory, the rate of dilepton production for this process,
within the EQPM can be written as
\begin{eqnarray} \label{dil_eqpm}
 \frac{dN}{d^4x d^4p} &=& \iint \frac{d^3\vec{p}_1}{(2\pi)^3} \frac{d^3\vec{p}_2}{(2\pi)^3}\,
 \frac{M_{\textrm{eff}}^2\,g^2 \,\sigma(M_{\textrm{eff}}^2)}{2 \omega_1 \omega_2}\nonumber\\
 && \times f_q(\vec{p}_1) f_q(\vec{p}_2) \delta^4(\tilde{p}-\tilde{p}_1-\tilde{p}_2).
\end{eqnarray}
Here, $\tilde{p}_{1,2} = (\omega_{1,2}, \vec{p}_{1,2})$ is the four-momentum of the quark, anti-quark
respectively, with $\omega_{1,2}$ being the corresponding modified single particle energy as given by 
Eq.~(\ref{E_modified}). When quark (anti-quark) masses are neglected, we can write
$\omega_{1,2} \approx |\vec{p}|$. 
Four-momentum of the dilepton is given as 
$\tilde{p} = (\omega_0 = \omega_1 + \omega_2, \vec{p} = \vec{p}_1 + \vec{p}_2)$. 
The quantity $M_{\textrm{eff}}^2 = (\omega_1 + \omega_2)^2 - (\vec{p}_1 + \vec{p}_2)^2$ 
represents the modified effective mass of 
the virtual photon in the interacting QCD medium. Keeping the terms up to linear order in $\delta \omega_q$,
we get~\cite{Naik:2021yph}
\begin{eqnarray}
 M_{\textrm{eff}}^2  &\approx& M^2 \left(1 + \frac{4\,\delta \omega_q\,(E_1 + E_2)}{M^2}\right),
\end{eqnarray}
where $M^2$ is the invariant mass of dilepton in the ultrarelativistic limit, $z_q \rightarrow 1$. In Eq.~\eqref{dil_eqpm}, the term $\sigma(M_{\textrm{eff}}^2)$ represents the cross-section for the $q \bar{q}$-annihilation process and $g$
is the degeneracy factor. 
In the Born approximation, we obtain $M_{\textrm{eff}}^2\,g^2 \,\sigma(M_{\textrm{eff}}^2)
= \frac{80\pi}{9} \alpha_e^2$~\cite{Alam:1996fd} for $N_f = 2$ and $N_c =3$, with $\alpha_e$ being the 
electromagnetic coupling constant.

We note that, in Eq.~\eqref{dil_eqpm}, $f_q(\vec{p}) \equiv f_q^0 + f_q^0 \bar{f}_q^0\delta f_q$ 
represents the quark (anti-quark) momentum distribution 
function away from equilibrium, with the form of equilibrium distribution function $f_q^0$ as given by Eq.~\eqref{QPM_f0}. The viscous modification to the distribution function (Eq.~(\ref{delta_f1}))
can be rewritten in terms of first order gradients of hydrodynamic quantities by employing 
Eq.~(\ref{I_order_ev}) and the form of $\delta f_q$ thus obtained is given below:
\begin{eqnarray}\label{delta_f_2}
 \delta f_q &=& \delta f_\pi + \delta f_\Pi \nonumber\\
 &=&\frac{\beta}{2 \beta_\pi (u \cdot \tilde{p})}
 \tilde{p}^\mu \tilde{p}^\nu \pi_{\mu\nu} + 
 \frac{\beta \Pi}{\beta_\Pi} \Big[\xi_1 - \xi_2 (u \cdot \tilde{p}) \Big],
\end{eqnarray}
where
\begin{eqnarray}  \label{coeff_xi}
 \xi_1 &=& \beta c_s^2 \frac{\partial \delta \omega_q}{\partial \beta} + \delta \omega_q,  \nonumber\\
 \xi_2 &=& \left( c_s^2 - \frac{1}{3} \right)  
 + \frac{\delta \omega_q}{3(u \cdot \tilde{p})^2 } \left[2(u \cdot \tilde{p})-\delta \omega_q\right].
\end{eqnarray}
We intend to calculate the spectra of dileptons with large invariant mass i.e., $M >> T$. 
Hence we can approximate the distribution function with the Maxwell-Boltzmann one, 
$f(\vec{p}) \approx z_q e^{-\omega/T}$.
Keeping the terms up to second order in momenta, we write the total dilepton production rate as~\cite{Bhatt:2011kx}
\begin{equation}  \label{dil_total}
 \frac{dN}{d^4x d^4p} = \frac{dN^{(0)}}{d^4x d^4p} + \frac{dN^{(\pi)}}{d^4x d^4p} 
 + \frac{dN^{(\Pi)}}{d^4x d^4p}.
\end{equation}
The ideal part of the rate, within the EQPM is obtained as~\cite{Chandra:2015rdz,Naik:2021yph}
\begin{eqnarray}
 \frac{dN^{(0)}}{d^4x d^4p} &=& z_q^2 \int \frac{d^3 \vec{p}_1}{(2\pi)^6}\,
 \frac{M_{\textrm{eff}}^2\,g^2\,\sigma(M_\textrm{eff}^2)}{2 \omega_1 \omega_2}\,
 e^{-(\omega_1 + \omega_2)/T} \nonumber\\
 && \times \delta(\omega_0 -\omega_1 -\omega_2) \nonumber \\
 &=& \frac{z_q^2}{2} \frac{M_{\textrm{eff}}^2\,g^2\,\sigma(M_{\textrm{eff}}^2)}{(2\pi)^5}e^{-\omega_0/T}.
\end{eqnarray}
The contribution to the thermal dilepton rate due to shear viscosity can be written as
\begin{eqnarray}
\frac{dN^{(\pi)}}{d^4x d^4p} &=& z_q^2 \iint \frac{d^3 \vec{p}_1}{(2\pi)^3}
\frac{d^3 \vec{p}_2}{(2\pi)^3}\,e^{-(\omega_1 + \omega_2)/T}    \nonumber\\
&& \times \left[ \frac{\beta}{\beta_\pi (u \cdot \tilde{p})}\right] 
\frac{M_{\textrm{eff}}^2\,g^2\,\sigma(M_\textrm{eff}^2)}{2 \omega_1 \omega_2}  \nonumber \\
&& \times \delta^4 (\tilde{p} - \tilde{p}_1 - \tilde{p}_2)\,\,
\tilde{p}_1^\mu \tilde{p}_1^\nu \pi_{\mu\nu}.
\end{eqnarray}
We note that, shear viscous correction to distribution function, $\delta f_\pi$ is qualitatively 
similar to the non-equilibrium correction obtained in Refs.~\cite{Bhalerao:2013aha,Naik:2021yph,Naik:2021yue}. 
Hence, by 
following the analysis of the above 
Refs., we obtain the contribution to the dilepton rate due to shear viscosity as
\begin{eqnarray}
 \frac{dN^{(\pi)}}{d^4x d^4p} &=& \frac{d N^{(0)}}{d^4 x d^4 p}
\Bigg\{\frac{\beta}{\beta_{\pi}} \frac{1 }{ 2|\vec{p}|^5 }
\Bigg[\frac{(u\cdot \tilde{p}) |\vec{p}|}{2}
\left(2|\vec{p}|^2-3 M_{\textrm{eff}}^2\right)\nonumber\\
&&+\frac{3}{4} M_{\textrm{eff}}^4  
\ln \left( \frac{(u\cdot \tilde{p}) + |\vec{p}|}{(u\cdot \tilde{p}) - |\vec{p}|} \right)\Bigg]
\tilde{p}^\mu \tilde{p}^\nu \pi_{\mu\nu}   \Bigg\},
\end{eqnarray}
where $|\vec{p}| = \sqrt{(u\cdot \tilde{p})^2 - M_{\textrm{eff}}^2}$.
Following the same analysis, we calculate the contribution to the dilepton rate due to bulk viscosity as
\begin{eqnarray}
 \frac{dN^{(\Pi)}}{d^4x d^4p} &=& \frac{dN^{(0)}}{d^4x d^4p} \frac{2\beta \Pi}{\beta_\Pi}
 \Bigg\{\beta c_s^2 \frac{\partial \delta\omega_q}{\partial\beta} 
 - \frac{2}{3}\delta\omega_q 
 - \left(c_s^2 -\frac{1}{3} \right)\frac{(u \cdot \tilde{p})}{2} \nonumber \\
 &&+ \frac{\delta\omega_q^2}{3} 
 \frac{1}{2 |\vec{p}|^5} \Bigg[\frac{(u\cdot \tilde{p}) |\vec{p}|}{2}
\left(2|\vec{p}|^2-3 M_{\textrm{eff}}^2\right)\nonumber\\
&&+\frac{3}{4} M_{\textrm{eff}}^4  
\ln \left( \frac{(u\cdot \tilde{p}) + |\vec{p}|}{(u\cdot \tilde{p}) 
- |\vec{p}|} \right)\Bigg]\Bigg\}.   
\end{eqnarray}
%
%
 %

We now determine the modification to the thermal photon emission rate due to shear and bulk viscosities.
We consider two major sources of thermal photons: \textit{Compton scattering}, $q(\bar{q}) g \rightarrow
q(\bar{q}) \gamma$ and \textit{$q \bar{q}$-annihilation}, $q \bar{q} \rightarrow g \gamma$.
We emphasize that, in the present work we only examine the hard contributions of thermal photon
emission. 
Following~\cite{Bhatt:2010cy,Dusling:2009bc,Wong:1995jf}, the total photon rate in the presence of dissipation within EQPM can be written as
\begin{eqnarray}
\omega_0 \frac{d N_\gamma}{d^3 p d^4 x} &=& \frac{5}{9} \frac{\alpha_{e} \alpha_{s}}{2 \pi^{2}}
T^2 f_q(\vec{p}) \left[ \ln \left(\frac{12\,(u\cdot \tilde{p})}{g^{2} T}\right)
 \!+\! \frac{C_{\textrm{ann}} \!+\! C_{\textrm{Comp}}}{2} \right]   \nonumber \\
&=& \omega_0 \frac{d N_\gamma^{(0)}}{d^3 p d^4 x} 
 + \omega_0 \frac{d N_\gamma^{(\pi)}}{d^3 p d^4 x} + \omega_0 \frac{d N_\gamma^{(\Pi)}}{d^3 p d^4 x},
\end{eqnarray}
where the constants take the values $C_{\textrm{ann}} = -1.91613$, $C_{\textrm{Comp}} = -0.41613$.
Here $\alpha_s$ denotes the strong coupling constant and $g = \sqrt{4\pi \alpha_s}$. 
The ideal contribution to the photon rate takes the following form: 
\begin{equation}
   \omega_0 \frac{d N_\gamma^{(0)}}{d^3 p d^4 x} = \frac{5}{9} \frac{\alpha_{e} \alpha_{s}}{2 \pi^{2}}
 T^{2}z_q^2 e^{-(u\cdot \tilde{p}) / T} \ln \left(\frac{3.7388\,(u\cdot \tilde{p})}{g^{2} T}\right). 
\end{equation}
Viscous contributions to photon rate are obtained as
\begin{eqnarray}
 \omega_0 \frac{d N_\gamma^{(\pi)}}{d^3 p d^4 x} &=& \omega_0 \frac{d N_\gamma^{(0)}}{d^3 p d^4 x}
 \left\{ \frac{\beta}{2 \beta_\pi (u \cdot \tilde{p})} \right\}, \\
 \omega_0 \frac{d N_\gamma^{(\Pi)}}{d^3 p d^4 x} &=& \omega_0 \frac{d N_\gamma^{(0)}}{d^3 p d^4 x}
 \left\{ \frac{\beta \Pi}{\beta_\Pi} \Big[\xi_1 - \xi_2 (u \cdot \tilde{p}) \Big]\right\},
\end{eqnarray}
where $\xi_1$ and $\xi_2$ are defined in Eq.~(\ref{coeff_xi}). 

Next, in order to do a comparative study with the standard hydrodynamic results, we calculate the thermal dilepton and photon emission rates by employing the viscous modified distribution function of the form: $f=f_0(1+\delta \phi)$, where $f_0 = e^{-E/T}$. The non-equilibrium correction is obtained from the $14-$moment  Grad's method and is shown below~\cite{Dusling:2007gi} : 
\begin{eqnarray} \label{grads_deltaf}
\delta \phi = \frac{p^\mu p^\nu}{2(\epsilon + P)T^2} \left(\pi_{\mu\nu} + \frac{2}{5}\Pi \Delta_{\mu\nu}\right).
\end{eqnarray}
The ideal part of the dilepton rate, in the absence of viscosity is well known and is given by~\cite{Vogt:2007zz}
\begin{equation}
   \frac{dN_0}{d^4x d^4p} = \frac{1}{2} \frac{M^2\,g^2\,\sigma(M^2)}{(2\pi)^5}e^{-(u\cdot p)/T}.
\end{equation}
The contribution to the dilepton rate due to shear and bulk viscous terms in $\delta \phi$ are calculated as~\cite{Dusling:2008xj,Bhatt:2011kx}
\begin{eqnarray}
\frac{dN_\pi}{d^4x d^4p} &=& \frac{2}{3}\left(\frac{p^\mu p^\nu}{2sT^3}\pi_{\mu\nu}\right)\frac{dN_0}{d^4x d^4p},\\
\frac{dN_\Pi}{d^4x d^4p} &=&  \frac{2}{5sT^3}\left(\frac{M^2}{12}g^{\alpha\beta}-\frac{1}{3}p^\alpha p^\beta\right)\Delta_{\alpha\beta}\Pi\frac{dN_0}{d^4x d^4p},\nonumber\\
\end{eqnarray}
respectively. Similarly, the ideal and viscous contributions to thermal photon rate are obtained as given below~\cite{Dusling:2009bc,Bhatt:2010cy} :
\begin{eqnarray}
   E \frac{d N_0}{d^3 p d^4 x} &=& \frac{5}{9} \frac{\alpha_{e} \alpha_{s}}{2 \pi^{2}}
 T^{2}e^{-(u\cdot p) / T} \ln \left(\frac{3.7388\,(u\cdot p)}{g^{2} T}\right), \\
 E \frac{d N_{\pi}}{d^3 p d^4 x} &=& E \frac{d N_{0}}{d^3 p d^4 x}
 \left\{ \frac{p^\mu p^\nu \pi_{\mu\nu}}{2(\epsilon + P)T^2} \right\}, \\
 E \frac{d N_{\Pi}}{d^3 p d^4 x} &=& E \frac{d N_{0}}{d^3 p d^4 x}
 \left\{\frac{2}{5}\Pi\frac{p^\mu p^\nu \Delta_{\mu\nu}}{2(\epsilon + P)T^2}\right\}.
\end{eqnarray}

In the next section, we proceed to calculate the thermal dilepton and photon spectra from heavy-ion collisions
by convoluting the rate expressions over the space-time evolution of the collisions along with the 
temperature profile and viscous evolution of the QGP obtained in section~\ref{Bjorken}.


\section{Particle spectra from an expanding QGP}
\label{yield}

Thermal dileptons and photons are produced from a thermalized QGP throughout its evolution. We calculate the emission
spectra of these particles over the entire evolution by considering Bj\"{o}rken's $1-$D model~\cite{Bjorken:1982qr}.
Within the model, four-dimensional volume element in the Minkowski space gets modified as
$d^4x=A_\bot \tau d\tau d\eta_s$. Here, $A_\bot = \pi R_A^2$ is the transverse area of the collision,
with $R_A = 1.2\,A^{1/3}$ fm being the radius of the colliding nuclei (for Au, $A=197$).
Now, we write the total thermal dilepton and photon yields in terms of transverse momentum $p_T$, 
invariant mass $M$ and rapidity 
$y$ of the particle produced
\begin{eqnarray}
 \frac{dN}{dM^2 d^2p_T dy} &=& A_\bot\,\int_{\tau_0}^{\tau_f}d\tau\,\tau \int_{-\infty}^\infty
 d\eta_s\,\chi(T,\eta_s)\left(\frac{1}{2} \frac{dN}{d^4 x d^4 p}\right) \nonumber\\  
\frac{dN}{d^2p_T dy} &=&A_\bot\,\int_{\tau_0}^{\tau_f}d\tau\,\tau \int_{-\infty}^\infty
 d\eta_s\,\left(\omega \frac{d N}{d^3 p d^4 x}\right),
 \end{eqnarray}
where the factor $\chi (T,\eta_s) = \left[1+\frac{2}{m_T}\cosh(y-\eta_s)\delta \omega_q\right]$ and $\tau_f$ is the time taken by the fireball to cool down to the critical temperature $T_c$.
Four-momentum of the dilepton can be parametrized as 
$\tilde{p}^\mu = (M_T\cosh y, p_T \cos \phi_p, p_T \sin\phi_p, M_T \sinh y)$, where
$M_T = \sqrt{p_T^2 + M_{\textrm{eff}}^2}$ is the medium modified transverse mass of dilepton. 
Now, for an expanding QGP within Bj\"{o}rken flow, we evaluate the factors appearing in particle rate 
as
\begin{align}
(u \cdot\tilde{p}) &= M_T \cosh(y-\eta_s),   \\
\tilde{p}^\mu \tilde{p}^\nu \pi_{\mu\nu} &= \pi \left[\frac{p_T^2}{2}
- M_T^2\sinh^2(y-\eta_s)\right], \\
\tilde{p}^\mu \tilde{p}^\nu \Delta_{\mu\nu} &= - p_T^2 - m_T^2 \sinh^2(y-\eta_s).
\end{align}
By noting the above expressions, we write the ideal contribution to thermal dilepton yields as 
\begin{eqnarray}
\frac{dN^{(0)}}{dM^2 d^2p_T dy} &=& \mathcal{Q}\int_{\tau_0}^{\tau_f}d\tau\,\tau z_q^2\,   
 \int_{-\infty}^\infty d\eta_s\,\chi(T,\eta_s)\nonumber \\ 
&&\times \textrm{Exp}\left(-\frac{M_T}{T} \cosh(y-\eta_s)\right).  \label{id_dil_yield} 
\end{eqnarray}
The viscous contributions to the dilepton yields are obtained as
\begin{widetext}
\begin{eqnarray}
 \frac{dN^{(\pi)}}{dM^2 d^2p_T dy} &=& \mathcal{Q}\int_{\tau_0}^{\tau_f}d\tau\,
 \tau z_q^2\,\frac{\beta \pi}{4\beta_\pi}
 \int_{-\infty}^\infty d\eta_s\,\chi(T,\eta_s)   
 \textrm{Exp}\left(-\frac{M_T}{T} \cosh(y-\eta_s)\right) 
 \frac{\left[p_T^2/2- M_T^2\sinh^2(y-\eta_s)\right]}
 {[M_T^2 \cosh^2(y-\eta_s) - M_{\textrm{eff}}^2]^{5/2}}   \nonumber\\
 &&\times \Bigg\{ M_T\cosh(y-\eta_s) \sqrt{M_T^2 \cosh^2(y-\eta_s) - M_{\textrm{eff}}^2}
 \Big(2M_T^2 \cosh^2(y-\eta_s) - 5M_{\textrm{eff}}^2 \Big) \nonumber \\
 &&+ \frac{3}{2}M_{\textrm{eff}}^4 \ln \left(
 \frac{M_T\cosh(y-\eta_s) + \sqrt{M_T^2 \cosh^2(y-\eta_s) - M_{\textrm{eff}}^2}}
 {M_T\cosh(y-\eta_s) - \sqrt{M_T^2 \cosh^2(y-\eta_s) - M_{\textrm{eff}}^2}}\right) \Bigg\},\\
 \frac{dN^{(\Pi)}}{dM^2 d^2p_T dy} &=&\mathcal{Q}\int_{\tau_0}^{\tau_f}d\tau\,
 \tau z_q^2 \frac{2\beta \Pi}{\beta_\Pi} 
 \int_{-\infty}^\infty d\eta_s\,\chi(T,\eta_s)   
 \textrm{Exp}\left(-\frac{M_T}{T} \cosh(y-\eta_s)\right) \nonumber \\
&&\times \Bigg\{\beta c_s^2 \frac{\partial \delta\omega_q}{\partial\beta} 
 - \frac{2}{3}\delta \omega_q - \frac{1}{2}\left(c_s^2 -\frac{1}{3}\right) 
 M_T \cosh (y-\eta_s) 
 + \frac{\delta \omega_q^2}{3} \frac{1}{4[M_T^2 \cosh^2(y-\eta_s) - M_{\textrm{eff}}^2]^{5/2}} \nonumber \\
 &&\times \Bigg[  M_T\cosh(y-\eta_s) \sqrt{M_T^2 \cosh^2(y-\eta_s) - M_{\textrm{eff}}^2}
 \Big(2M_T^2 \cosh^2(y-\eta_s) - 5M_{\textrm{eff}}^2 \Big) \nonumber \\
 &&+ \frac{3}{2}M_{\textrm{eff}}^4 \ln \left(
 \frac{M_T\cosh(y-\eta_s) + \sqrt{M_T^2 \cosh^2(y-\eta_s) - M_{\textrm{eff}}^2}}
 {M_T\cosh(y-\eta_s) - \sqrt{M_T^2 \cosh^2(y-\eta_s) - M_{\textrm{eff}}^2}}\right)\Bigg]\Bigg\},
\end{eqnarray}
\end{widetext}
where $\mathcal{Q} = \frac{A_\bot}{4 (2\pi)^5} \frac{80 \pi}{9}\alpha_e^2$. From 
Eq.~(\ref{dil_total}), we note the total dilepton yield to be 
\begin{equation}
 \frac{dN}{dM^2 d^2p_T dy} = \frac{dN^{(0)}}{dM^2 d^2p_T dy} +
 \frac{dN^{(\pi)}}{dM^2 d^2p_T dy} + \frac{dN^{(\Pi)}}{dM^2 d^2p_T dy}. \nonumber
\end{equation}
The thermal photon yield is obtained as 
\begin{equation}
 \frac{dN}{d^2p_T dy} = \frac{dN^{(0)}}{d^2p_T dy} 
 + \frac{dN^{(\pi)}}{d^2p_T dy} + \frac{dN^{(\Pi)}}{d^2p_T dy}.
\end{equation}
Noting the photon energy to be $(u \cdot \tilde{p}) = p_T \cosh(y-\eta_s)$, 
the ideal part of photon yield is obtained as
\begin{eqnarray}\label{ideal_ph_yield}
 \frac{dN^{(0)}}{d^2p_T dy} &=& \mathcal{C} \int_{\tau_0}^{\tau_f}d\tau\,\tau z_q^2 T^{2} 
 \int_{-\infty}^\infty d\eta_s  \textrm{Exp}\left(-\frac{p_T}{T}\cosh(y-\eta_s) \right) 
 \nonumber \\
&&\times \ln \left(\frac{3.7388\,p_T \cosh(y-\eta_s)}{g^{2} T}\right). \label{id_ph_yield}
\end{eqnarray}
Viscous contributions to photon yield are obtained as
\begin{eqnarray}
\frac{dN^{(\pi)}}{d^2p_T dy} &=& \mathcal{C} \int_{\tau_0}^{\tau_f}d\tau\,\tau z_q^2 T^{2} 
 \int_{-\infty}^\infty d\eta_s  \textrm{Exp}\left(-\frac{p_T}{T}\cosh(y-\eta_s) \right) 
 \nonumber \\
&&\times \frac{\beta}{2\beta_\pi} \frac{1}{p_T \cosh(y-\eta_s)} 
\ln \left(\frac{3.7388\,p_T \cosh(y-\eta_s)}{g^{2} T}\right) \nonumber \\
&&\times \pi \left[\frac{p_T^2}{2}
- M_T^2\sinh^2(y-\eta_s)\right] \\
\frac{dN^{(\Pi)}}{d^2p_T dy} &=& \mathcal{C} \,\int_{\tau_0}^{\tau_f}d\tau\,\tau z_q^2 T^{2} 
 \int_{-\infty}^\infty d\eta_s  \textrm{Exp}\left(-\frac{p_T}{T}\cosh(y-\eta_s) \right) 
 \nonumber \\
&&\times \frac{\beta \Pi}{\beta_\Pi} \left[\xi_1 - \xi_2 p_T \cosh(y-\eta_s) \right]\nonumber\\
&&\times \ln \left(\frac{3.7388\,p_T \cosh(y-\eta_s)}{g^{2} T}\right), 
\end{eqnarray}
where $\mathcal{C} = \frac{5A_\bot}{9} \frac{\alpha_{e} \alpha_{s}}{2 \pi^{2}}$.

Now, we write the expressions for thermal dilepton and photon yields in the standard hydrodynamics with the non-equilibrium corrections given by Eq.~\eqref{grads_deltaf}.
The ideal dilepton yield is obtained by taking the limit $z_q \rightarrow 1$ and $M_{eff} \rightarrow M$ in Eq.~\eqref{id_dil_yield}.
The viscous corrections to the dilepton yields are given by 
\begin{eqnarray}
\frac{dN_{\pi}}{dM^2 d^2p_T dy} &=& \mathcal{Q}\int_{\tau_0}^{\tau_f}d\tau\,\tau   
 \int_{-\infty}^\infty d\eta_s
 \textrm{Exp}\left(-\frac{m_T}{T} \cosh(y-\eta_s)\right)\nonumber \\
 &&\times\frac{\pi}{3sT^3} \left[\frac{p_T^2}{2}- m_T^2\sinh^2(y-\eta_s)\right], \nonumber\\
 \frac{dN_{\Pi}}{dM^2 d^2p_T dy} &=& \mathcal{Q}\int_{\tau_0}^{\tau_f}d\tau\,\tau  
 \int_{-\infty}^\infty d\eta_s\textrm{Exp}\left(-\frac{m_T}{T} \cosh(y-\eta_s)\right) \nonumber\\
&&\times\frac{2\Pi}{5sT^3}\left[\frac{M^2}{4}+\frac{1}{3}\Bigg(p_T^2 + m_T^2\sinh^2(y-\eta_s)\Bigg)\right]. \nonumber \\
\end{eqnarray}
The viscous corrections to the thermal photon yields are obtained as
\begin{eqnarray}
\frac{dN_{\pi}}{d^2p_T dy} &=& \mathcal{C} \,\int_{\tau_0}^{\tau_f}d\tau\,\tau T^{2} 
 \int_{-\infty}^\infty d\eta_s  \textrm{Exp}\left(-\frac{p_T}{T}\cosh(y-\eta_s) \right) 
 \nonumber \\
&&\times 
\ln \left(\frac{3.7388\,p_T \cosh(y-\eta_s)}{g^{2} T}\right)\nonumber \\
&& \times  \frac{\pi}{2sT^3}\left[\frac{p_T^2}{2}- p_T^2\sinh^2(y-\eta_s)\right], \nonumber\\
\frac{dN^{(\Pi)}}{d^2p_T dy} &=& \mathcal{C} \,\int_{\tau_0}^{\tau_f}d\tau\,\tau T^{2} 
 \int_{-\infty}^\infty d\eta_s  \textrm{Exp}\left(-\frac{p_T}{T}\cosh(y-\eta_s) \right) 
 \nonumber \\
&&\times \ln \left(\frac{3.7388\,p_T \cosh(y-\eta_s)}{g^{2} T}\right) \nonumber\\
&&\times \frac{\Pi}{5(\epsilon + P)T^2}\Big[- p_T^2 - m_T^2 \sinh^2(y-\eta_s)\Big].
\end{eqnarray}


\section{Results and Discussion}
\label{results}

In this section, we present the particle spectra in the presence of viscous corrections 
for longitudinal expansion 
of QGP medium. The temperature and viscous profiles ($T(\tau), \pi(\tau)$ and $\Pi(\tau)$) have been determined in section~\ref{Bjorken}
by numerically solving the hydrodynamic equations, 
Eqs.~\eqref{energy_evolution}-\eqref{Pi_evolution}. 
In order to obtain the dilepton and photon spectra from QGP phase, we evolve the system till $T_c$ and in this work, we fix $T_c = 0.17$ GeV. The value of $\tau_f$ is determined by solving the equation : $T(\tau)==T_c$.
Also, the spectra is calculated for the midrapidity region of the particles, $i.e.,\,y=0$. 

\begin{figure}[t!]
 \begin{center}
  \includegraphics[width=\linewidth]{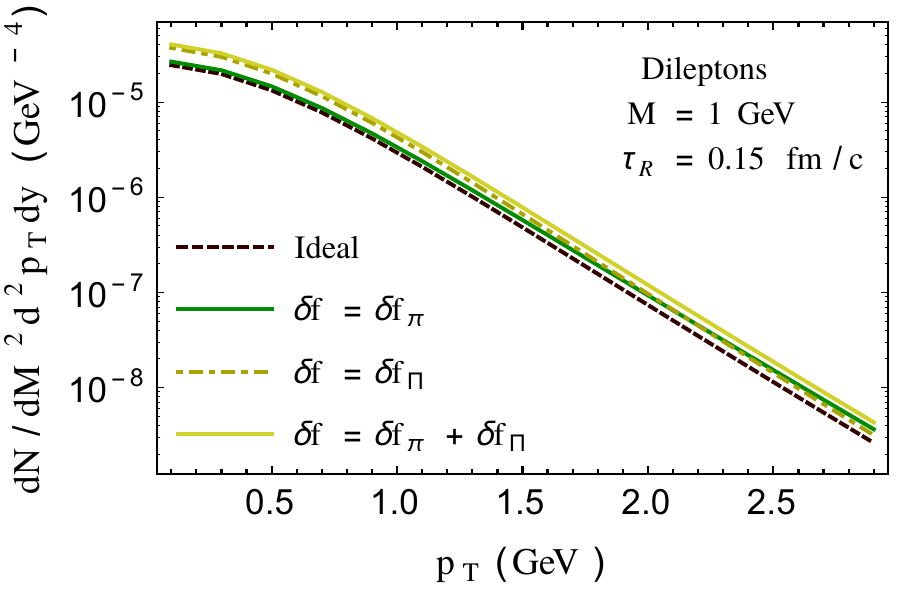}
 \end{center}
 \vspace{-0.8cm}
 \caption{Thermal dilepton yields in the presence of viscous corrections  corresponding to $\tau_R = 0.15$ fm/c and for $M=1$ GeV.  }
 \label{dil1}
\end{figure}
\begin{figure}[t!]
 \begin{center}
  \includegraphics[width=\linewidth]{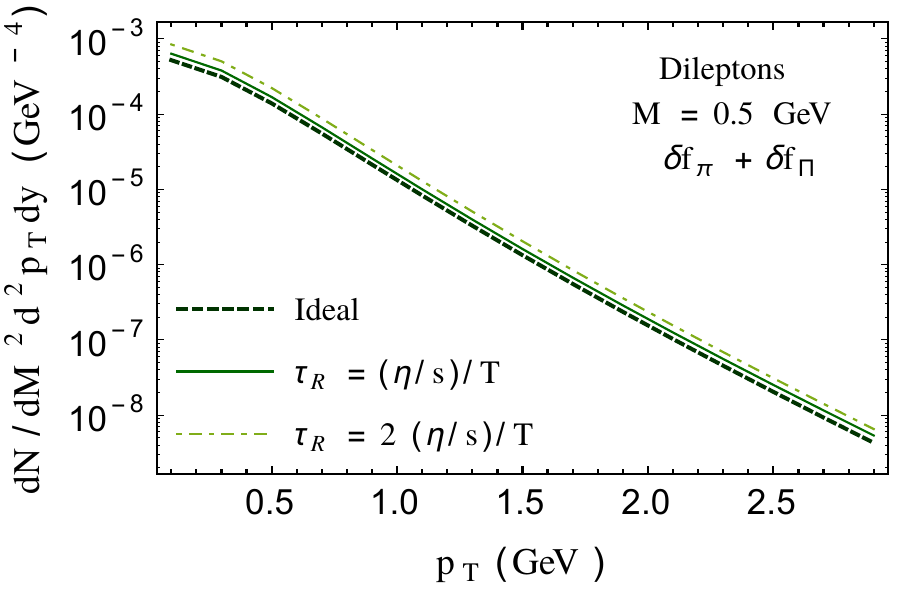}
 \end{center}
 \vspace{-0.8cm}
 \caption{Dilepton spectra in the presence of viscous corrections by varying $\tau_R$ for $M=1$ GeV}
 \label{dil2}
\end{figure}
\begin{figure}[t!]
 \begin{center}
  \includegraphics[width=\linewidth]{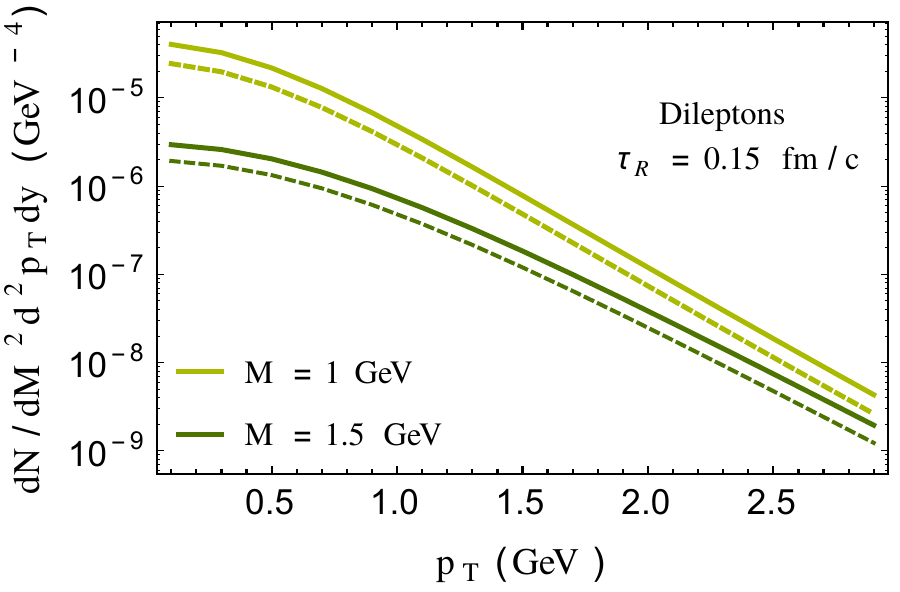}
 \end{center}
 \vspace{-0.8cm}
 \caption{Comparison of dilepton spectra for different $M$ values with $\tau_R = 0.15$ fm/c. The solid lines represent the total yields and dashed lines correspond to $\delta f=0$ case.}
 \label{dil3}
\end{figure}

We now study the effect of viscosity on particle spectra by comparing the total spectra with the 
ideal case ($\delta f = 0$). The ideal 
dilepton 
and photon spectra is calculated by integrating the ideal contribution
to the yields, Eqs.~\eqref{id_dil_yield} and \eqref{id_ph_yield} along with
the ideal Bj\"{o}rken evolution, $T_i(\tau) = T_0(\tau_0/\tau)^{1/3}$. Firstly, we analyze the impact of both shear and bulk viscous corrections to the distribution function on thermal dilepton yield. In Fig.~\ref{dil1},
we show thermal dilepton spectra in the presence of dissipation for a constant value of relaxation time, $\tau_R = 0.15$ fm/c. The yields are plotted for invariant mass  $M=1$ GeV. The dashed curve denotes the ideal case. We first consider the case, $ \delta f=\delta f_\pi$, which is computed by taking $\delta f_\Pi = 0$ and $\Pi = 0$ in the analysis. We observe that shear viscosity enhances the dilepton spectra, especially at high $p_T$.  However, this increment is observed to be marginal when compared to the yield in the presence of bulk viscosity. 
The yield for $\delta f_\Pi$ correction is plotted by switching off the shear contributions in Eqs.~\eqref{delta_f_2} and \eqref{energy_evolution}-\eqref{Pi_evolution}. Fig.~\ref{dil1} shows that the 
effect of bulk viscosity enhances the dilepton yield throughout the $p_T$ regime. Enhancement is observed to be maximum at small $p_T$ and minimum at large $p_T$. 
Moreover, maximum increment in yield is observed when 
both shear and bulk corrections ($\delta f_\pi + \delta f_\Pi$) are taken into account. 

Next, we compare the thermal dilepton yields corresponding to $M=1$ GeV for different temperature dependent $\tau_R$ in Fig.~\ref{dil2}. We take both $\delta f_\pi$ and $\delta f_\Pi$ terms in distribution function for this comparison. 
It can be seen that the dilepton yield increases with the increase in magnitude of $\tau_R$.
The maximum enhancement in the yield is observed for $\tau_R = 2(\eta/s)/T$ and 
minimum for $\tau_R = (\eta/s)/T$. In Fig.~\ref{dil3}, we study the impact of viscosities 
on dilepton yield by varying the value of $M$. We plot the total yields for $M=1,1.5$ GeV. 
The dashed lines indicate the corresponding yields for $\delta f=0$. 
We observe that there is an overall decrement in the yield as the value of $M$ increases.
Also, note that enhancement to the yield due to dissipation decreases with large $M$. 
Over the entire $p_T$ range, enhancement due to  viscosity is more for $M=1$ GeV. 

\begin{figure}[t!]
 \begin{center}
  \includegraphics[width=\linewidth]{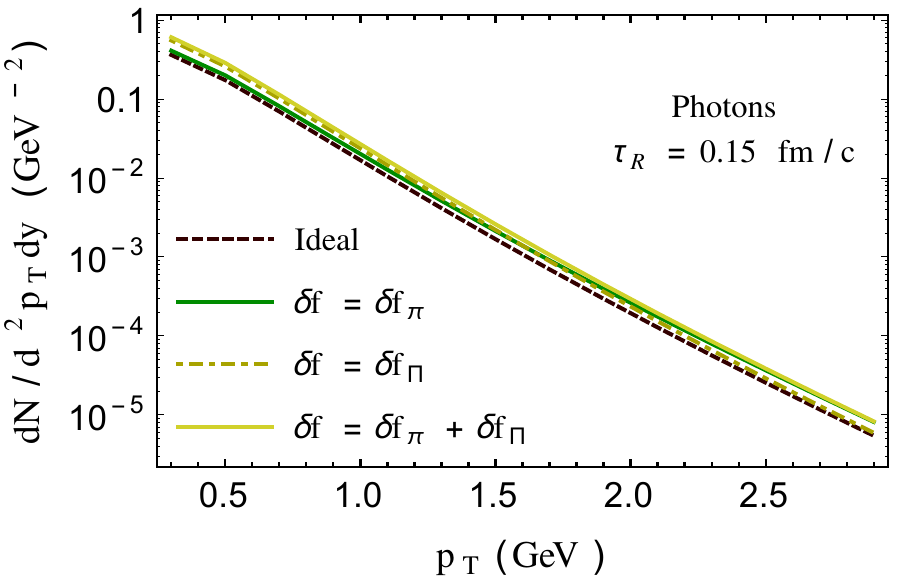}
 \end{center}
 \vspace{-0.8cm}
 \caption{Thermal photon yield in the presence of dissipative corrections for constant relaxation time, $\tau_R = 0.15$ fm/c. }
 \label{ph1}
\end{figure}

\begin{figure}[t]
 \begin{center}
  \includegraphics[width=\linewidth]{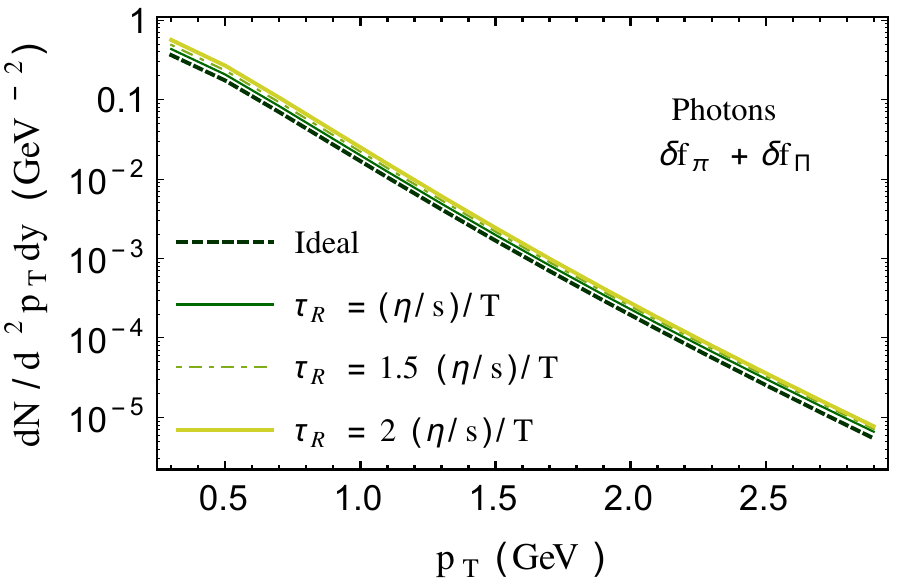}
 \end{center}
 \vspace{-0.8cm}
 \caption{Comparison of thermal photon yields in the presence of total viscous correction ($\delta f_\pi + \delta f_\Pi$) by varyiing the relaxation time.}
 \label{ph2}
\end{figure}

Similarly, we analyse the effect of dissipation on thermal photon yield in Figs.~\ref{ph1} and \ref{ph2}. 
Impact of viscous terms on photon yield is studied with $\tau_R=0.15$ fm/c in Fig.~\ref{ph1}.
As observed for dilepton (in Fig.~\ref{dil1}), maximum enhancement to the yield due to bulk viscosity ($\delta f_\Pi$) is observed at small $p_T$, where as increment due to shear correction ($\delta f_\pi$) is significant at high $p_T$.
The combined effect of both ($\delta f_\pi + \delta f_\Pi$) leads to the maximum enhancement over the entire $p_T$. Moreover, we note that the yield in the presence of bulk viscous correction crosses the $\delta f_\pi$ curve around $p_T \sim 1.7$ GeV, where as in Fig.~\ref{dil1} (for dileptons), this cross-over was observed around $p_T \sim 2$ GeV.
Fig.~\ref{ph2} exhibits a comparison between thermal photon yields for different $\tau_R$. It is observed that enhancement in the yield is seen to be marginal with $\tau_R = (\eta/s)/T$ and yield is greatest with $\tau_R = 2(\eta/s)/T$.

\begin{figure}[t]
 \begin{center}
  \includegraphics[width=\linewidth]{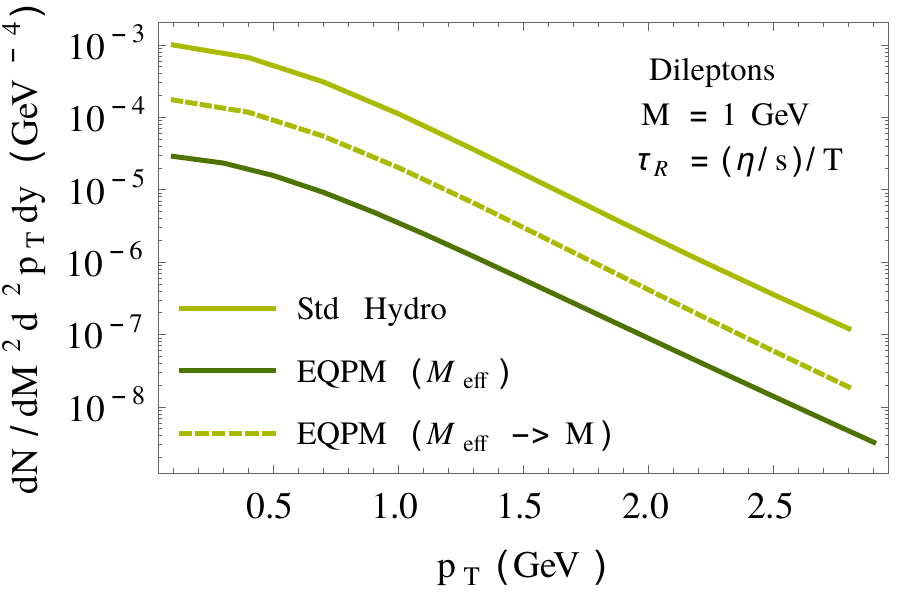}
 \end{center}
 \vspace{-0.8cm}
 \caption{Comparison of thermal dilepton yields obtained within EQPM with the yields calculated using the standard hydrodynamic framework in Ref.~\cite{Bhalerao:2013aha}. }
 \label{dil_compare}
\end{figure}

\begin{figure}[t]
 \begin{center}
  \includegraphics[width=\linewidth]{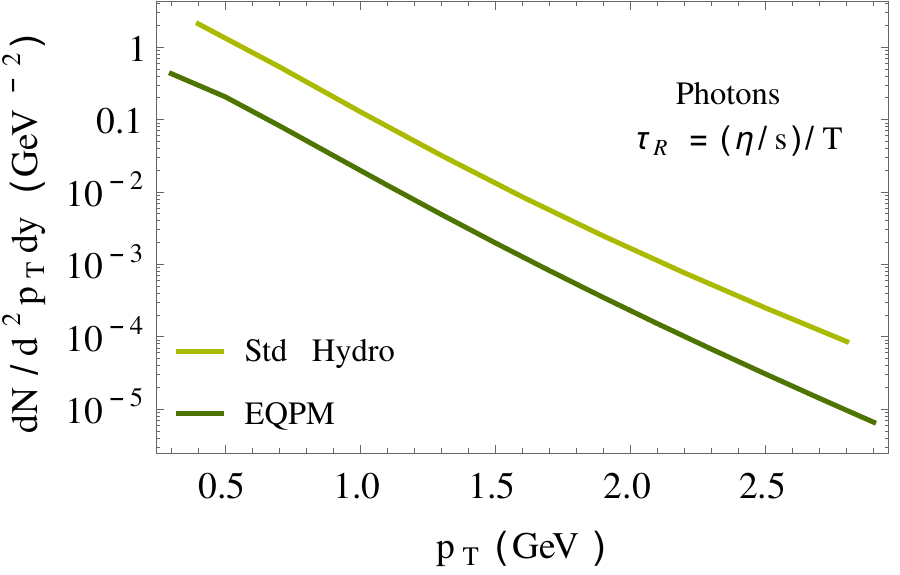}
 \end{center}
 \vspace{-0.8cm}
 \caption{Comparison of thermal photon yields obtained within EQPM with the yields calculated using the standard hydrodynamic framework in Ref.~\cite{Bhalerao:2013aha}. }
 \label{ph_compare}
\end{figure}

Finally, in Figs.~\ref{dil_compare} and \ref{ph_compare}, we compare the thermal dilepton and photon yields obtained within this new second order hydrodynamics with that of a standard one given by Eqs.~\eqref{energy_evolution}, \eqref{shear_std} and \eqref{bulk_std}. The yields are plotted for $\tau_R = (\eta/s)/T$ and with invariant mass $M=1$ GeV (for dileptons). It is clear that the spectra of dileptons and photons depend strongly on the interaction effects in the medium. The presence of
these interaction terms in the yield expressions within EQPM suppresses the particle production throughout the $p_T$ regime. This is inline with the results of Ref.~\cite{Chandra:2016dwy}.
In Fig.~\ref{dil_compare}, we also plot the dilepton yield within EQPM by neglecting the interaction effects on invariant mass of dilepton i.e., in the limit $M_{eff} \rightarrow M$. It can be seen that the dilepton production yield increases in this limit as shown by the results of Ref.~\cite{Naik:2020jfc}. 



\section{Summary}
\label{summary}
We have employed the recently developed second order dissipative hydrodynamics estimated within the effective 
fugacity quasiparticle model to study the thermal dilepton and photon production from QGP.
In this study, we have used the non-equilibrium 
quark-antiquark distribution functions, up to first order in momenta, determined using the iterative 
CE type expansion of effective Boltzmann equation in the relaxation time approximation (RTA). 
The second order viscous hydrodynamic equations were solved for different temperature dependent relaxation times ($\tau_R$) within the $1-$D boost invariant expansion of QGP and the evolution
has been compared with that obtained for a temperature independent constant value $\tau_R=0.15$ fm/c. It must be emphasized that, the choice of this constant value is 
arbitrary.
We have incorporated the effect of shear and bulk viscosity coefficients through their respective relaxation times $\tau_\pi$ and $\tau_\Pi$, ensuring $\tau_\pi=\tau_\Pi$ as demanded by RTA. 
The impact of shear and bulk viscous pressures was found to increase with increment
in $\tau_R$.
By looking into the dynamical pressure anisotropy with various relaxation times, we found that cavitation scenarios can be present in the medium for small values of $\tau_R$. 
Moreover, 
we obtained the limiting values of relaxation times by looking into fireball reheating scenarios. 
Our analysis indicate that relaxation time should not increase the constant value $0.5$ fm/c and temperature dependent value 
$8(1/4\pi)/T$.

Further, using this causal hydrodynamic model, 
we explored the thermal dilepton and photon yields from prominent production sources 
in the presence of viscous modified distribution functions within 
longitudinal expansion of the QGP. This was done after calculating the thermal particle spectra by including the  viscous modified single particle distribution functions. 
The effect of both shear and bulk corrections under this formalism 
is to enhance the dilepton and photon spectra. 
We also analyzed the impact of dissipation on yields by varying the value of invariant mass. 
Thermal dilepton and photon yields were studied for different temperature dependent $\tau_R$ and 
we found that maximum enhancement is observed for the largest value of $\tau_R$ considered.  
Our results also indicate that thermal dilepton and photon spectra in the presence of first order dissipative terms 
in the distribution function does not exhibit large corrections from the ideal case and is well behaved compared to 
the results using Grad's method~\cite{Bhatt:2011kx, Bhalerao:2013pza}.
We also did a comparative study of the spectra calculated within EQPM hydrodynamic framework with that obtained in the absence of mean field terms, by employing a standard relativistic hydrodynamics. Our study indicates that the presence of interaction terms in the yield expressions suppress the thermal dilepton and photon spectra throughout the $p_T$ range. 
In future, we would like to study thermal dilepton and photon production by 
employing this causal second order hydrodynamic framework in the presence of Chapman-Enskog type 
viscous corrections, up to second order in gradients~\cite{Bhadury:2020ivo}. 
Further, it will be interesting to do a quantitative analysis with ($2+1$)-D hydrodynamics by varying the initial conditions.
We leave these aspects for future study.

\section*{Acknowledgements}
L. J. N. acknowledges the  Department of Science and Technology, Government of India for the INSPIRE Fellowship. We thank the anonymous referees of this article for comments which led to the improvement of quality of the manuscript.
\appendix
\section{Second Order Transport Coefficients} \label{A}

The second order transport coefficients appearing in the viscous evolution Eqs.~\eqref{shear evolution def hydro} and \eqref{bulk evolution def hydro} are obtained as \cite{Bhadury:2020ivo}
\begin{widetext}
\begin{align}
 \delta_{\pi\pi}&=\dfrac{5}{3}+\dfrac{\beta}{\beta_{\pi}}\sum_{k=q,g}\bigg[\dfrac{7}{3}\Tilde{J}^{(3)}_{k~63}+{\delta{\omega}_k}\Big(\dfrac{7}{3}\Tilde{L}^{(3)}_{k~63}-\dfrac{7}{6}\xi_k+\dfrac{1}{2}\Gamma_k\Big)\bigg],\label{delta pipi}\\
\tau_{\pi\pi}&= 2 + \frac{\beta}{\beta_{\pi}} \sum_{k=q,g} \bigg[4\Big(\Tilde{J}^{(3)}_{k~63}+\delta\omega_k\Tilde{L}^{(3)}_{k~63}\Big)-{2\delta\omega_k }\xi_k\bigg],\label{tau pipi}   
\end{align}
\end{widetext}
\begin{widetext}
\begin{align}
\lambda_{\pi\Pi}&= \dfrac{\beta c_s^2}{\beta_{\Pi}}\sum_{k=q,g}\bigg[\Tilde{J}^{(1)}_{k~42}+\Tilde{J}^{(0)}_{k~31}+\delta\omega_k\Big(\Tilde{L}^{(1)}_{k~42}-\Tilde{J}^{(0)}_{k~21}+\Tilde{L}^{(0)}_{k~31}-\delta\omega_k\Tilde{L}^{(0)}_{k~21}\Big) +\beta\dfrac{\partial\delta\omega_k}{\partial\beta}\Big(2\xi_k+\Gamma_k\nonumber\\
&~~-2\delta\omega_k\Tilde{L}^{(1)}_{k~42}\Big)\!\bigg]+\dfrac{\beta}{\beta_{\Pi}}\sum_{k=q,g}\!\!\bigg[\dfrac{14}{3}\Tilde{J}^{(3)}_{k~63}+\dfrac{10}{3}\Tilde{J}^{(1)}_{k~42}+\delta\omega_k\Big(\dfrac{14}{3}\Tilde{L}^{(3)}_{k~63}+\dfrac{10}{3}\Tilde{L}^{(1)}_{k~42}-\dfrac{7}{3}\xi_k+\Gamma_k\Big)\!\bigg],\label{lambda piPi}\\
\lambda_{\Pi\pi} &= \dfrac{\beta}{3\beta_{\pi}}\sum_{k=q,g}\bigg[7\Tilde{J}^{(3)}_{k~63}+2\Tilde{J}^{(2)}_{k~52}+\delta\omega_k\Big(7\Tilde{L}^{(3)}_{k~63}+2\Tilde{L}^{(2)}_{k~52}-2\xi_k\Big)\bigg]-c_s^2,\label{lambda Pipi}\\
\delta_{\Pi\Pi}&=\frac{\beta}{\beta_{\Pi}} \!\! \sum_{k=q,g} \!\! \bigg[ \!\!-\! \frac{5}{9}\lambda_{0k} \!-\! \delta\omega_k\lambda_{1k} \!+\! \!\left(\!\frac{\partial\delta\omega_k}{\partial\beta}\!\right)\!\lambda_{2k} \!-\! (\delta\omega_k)^2 \lambda_{3k}\!+\! \delta\omega_k \!\left(\!\frac{\partial\delta\omega_k}{\partial\beta}\!\right)\!\! \lambda_{4k} \!-\! \left(\!\frac{\partial\delta\omega_k}{\partial\beta}\!\right)^2 \!\!\!\lambda_{5k}\!\bigg] \!-\! c_s^2, \label{delta PiPi}  
\end{align}
where,
\begin{align}
\xi_k &=~ \Tilde{J}^{(2)}_{k~42}+\delta\omega_k\Tilde{L}^{(2)}_{k~42},\\
\Gamma_k &=~ \Tilde{J}^{(0)}_{k~21}-\beta \Tilde{M}^{(1)}_{k~42} +\delta\omega_k\Big(\Tilde{L}^{(0)}_{k~21}-\Tilde{J}^{(1)}_{k~21}-\beta \Tilde{N}^{(1)}_{k~42} \Big),\\
\lambda_{1k} &=~ \Big(\dfrac{8}{3}\Tilde{J}^{(0)}_{k~21}-\beta \Tilde{M}^{(0)}_{k~31}\Big)c_s^2+\dfrac{25}{9}\Tilde{J}^{(2)}_{k~42}-\dfrac{5}{3}\Tilde{J}^{(1)}_{k~31}-\dfrac{5}{3}\beta\Tilde{M}^{(1)}_{k~42},\\
\lambda_{2k} &=~ \dfrac{5}{3}\Big(\Tilde{J}^{(1)}_{k~31}+\Tilde{J}^{(2)}_{k~42}+\beta\Tilde{M}^{(1)}_{k~42}-\Tilde{L}^{(1)}_{k~42}\Big)\beta c_s^2+\Tilde{M}^{(0)}_{k~31}\beta^2(c_s^2)^2,\\
\lambda_{3k} &=~ \dfrac{5}{3}\Tilde{J}^{(1)}_{k~21}-\beta\Tilde{M}^{(0)}_{k~21}+\Big(\dfrac{8}{3}\Tilde{L}^{(0)}_{k~21}-\beta\Tilde{N}^{(0)}_{k~31}\Big)c_s^2+\dfrac{25}{9}\Tilde{L}^{(2)}_{k~42}-\dfrac{5}{3}\Tilde{L}^{(1)}_{k~31}-\dfrac{5}{3}\beta\Tilde{N}^{(1)}_{k~42},\\
\lambda_{4k} &=~ \Big(\dfrac{1}{3}\Tilde{J}^{(1)}_{k~21}+2\beta\Tilde{M}^{(0)}_{k~21}\Big)\beta c_s^2+\Tilde{N}^{(0)}_{k~31}\beta^2 (c_s^2)^2+\dfrac{5}{3}\Big(\dfrac{3}{5}\Tilde{L}^{(0)}_{k~21}+\Tilde{L}^{(1)}_{k~31}+\Tilde{L}^{(2)}_{k~42}+\beta\Tilde{N}^{(1)}_{k~42}\Big)\beta c_s^2,\\
\lambda_{5k} &=~ \Big(\Tilde{J}^{(1)}_{k~21}-\beta\Tilde{M}^{(0)}_{k~21}+\beta^{-1}\Tilde{L}^{(0)}_{k~31}\Big)\beta^2 (c_s^2)^2.
\end{align}

Thermodynamic integrals appearing in the second order transport coefficients are as shown below
\begin{align}
\Tilde{J}^{(r)}_{k~nm}&=\dfrac{g_{k}}{2\pi^2}\frac{(-1)^m}{(2m+1)!!}\int_{0}^\infty{d\mid\vec{\Tilde{p}}_k\mid}~{\big(u.\Tilde{p}_k\big)^{n-2m-r-1}}\big(\mid\vec{\Tilde{p}}_k\mid\big)^{2m+2}f^0_k\bar{f}^0_k, \label{Jnmr}\\
\Tilde{L}^{(r)}_{k~nm}&=\dfrac{g_{k}}{2\pi^2}\frac{(-1)^m}{(2m+1)!!}\int_{0}^\infty{d\mid\vec{\Tilde{p}}_k\mid}~\dfrac{\big(u.\Tilde{p}_k\big)^{n-2m-r-1}}{E_k}\big(\mid\vec{\Tilde{p}}_k\mid\big)^{2m+2}f^0_k\bar{f}^0_k.\label{Lmnr}   \\
\Tilde{M}^{(r)}_{k~nm} =&~ \dfrac{g_{k}}{2\pi^2}\frac{(-1)^m}{(2m+1)!!}\int_{0}^\infty{d\mid\vec{\Tilde{p}}_k\mid}~{\big(u.\Tilde{p}_k\big)^{n-2m-r-1}}\big(\mid\vec{\Tilde{p}}_k\mid\big)^{2m+2}(\bar{f}^0_k-af^0_k)f^0_k\bar{f}^0_k, \label{Mmnr}\\
\Tilde{N}^{(r)}_{k~nm} =&~ \dfrac{g_{k}}{2\pi^2}\frac{(-1)^m}{(2m+1)!!}\int_{0}^\infty{d\mid\vec{\Tilde{p}}_k\mid}~\dfrac{{\big(u.\Tilde{p}_k\big)^{n-2m-r-1}}}{E_k}\big(\mid\vec{\Tilde{p}}_k\mid\big)^{2m+2}(\bar{f}^0_k-af^0_k)f^0_k\bar{f}^0_k. \label{Nmnr}
\end{align}
\end{widetext}

We note that, 
these thermodynamic integrals can be 
approximated and expressed in terms of 
polylogrithmic functions
of order $n$ and argument $a$, PolyLog$[n,a]$ as shown in Ref.~\cite{Bhadury:2020ivo}.


\bibliography{references}{}

\end{document}